# Aperiodic metalenses with identical nanocylinders: positional phase control and passive chromatic aberration suppression

Ivan Moreno, J. Carlos Basilio-Ortiz

Unidad Académica de Ciencia y Tecnología de la Luz y la Materia, Universidad Autonoma de Zacatecas, 98060 Zacatecas, Mexico
*e-mail:* imorenoh@uaz.edu.mx

**Abstract:** Conventional metalenses control light by varying meta-atom geometry, a design strategy that inherently couples phase modulation to structural dimensions, contributing to chromatic dispersion. Here, we introduce a new design paradigm in which phase control is decoupled from meta-atom geometry. We present an aperiodic metalens architecture composed exclusively of structurally identical, polarization-insensitive dielectric nanorods. In this framework, full 2π phase coverage is achieved solely through local periodicity modulation, that is, by varying the spacing between adjacent nanocylinders. We theoretically demonstrate that this geometric invariance yields a linear dependence of the effective refractive index on the local fill factor, which naturally approaches the dispersive condition required for achromatic focusing. The resulting architecture provides a deterministic, library-free, and analytically tractable approach to metalens design, in which positional degrees of freedom replace geometric variation as the phase-encoding mechanism. An extensive set of full-wave numerical simulations demonstrates the viability of the concept for visible-light focusing, showing that the aperiodic architecture can passively suppress chromatic aberration while maintaining near-diffraction-limited focusing. These results establish geometry-invariant positional phase control as a distinct design framework and provide a simple, reproducible baseline for further dispersion-engineered flat optics.

## 1. Introduction

Driven by the relentless demand for miniaturized and integrated optical systems, metalenses have emerged as a paradigm shift, offering the functionality of conventional bulk optics within an ultra-compact, planar form factor [1,2]. These devices consist of a metasurface, a dense array of subwavelength nano-elements or meta-atoms, that collectively sculpt the phase profile of incident light to achieve precise wavefront shaping [3,4]. A pivotal evolution in the field was the transition from early metallic designs, which suffered from significant optical losses, to transparent dielectric building blocks. This breakthrough dramatically improved optical efficiency, positioning metalenses as a transformative solution for next-generation imaging applications ranging from consumer electronics to medical diagnostics.

Metalenses are constructed from spatially arranged subwavelength meta-atoms, each engineered to impart a specific local phase shift and thereby collectively achieve a target phase profile. This phase control is achieved through different phase-delay mechanisms, which are established by spatially modulating the geometric parameters of the meta-atoms, such as shape, size, and orientation. The primary phase shift mechanisms utilized are resonant phase, geometric (Pancharatnam–Berry) phase, and propagation phase [4]. Our work focuses on the propagation phase, a mechanism prized for its polarization-insensitive behavior [5]. However, propagation-phase metalenses inherently couple the local phase response to the meta-atom geometry, introducing a nonlinear dispersive effective index that varies across the aperture.

Therefore, efforts to extend propagation-phase designs toward broadband chromatic correction have increasingly relied on advanced strategies, including inverse design, generative methods, free-form unit cells, and large geometry libraries [6–9].

Conventionally, achieving full phase coverage via the propagation phase mechanism requires varying the cross-sectional dimensions of the nano-elements, a common approach being the modulation of cylindrical nanorod diameters [10]. This work, however, introduces a fundamental departure from this size-phase dependency. Instead of varying the nano-element dimensions, we maintain structurally identical nanocylinders and achieve the desired phase profile by precisely engineering the local periodicity between them. This conceptual shift, schematically illustrated in Figure 1, effectively decouples the phase response from the meta-atom's geometry.

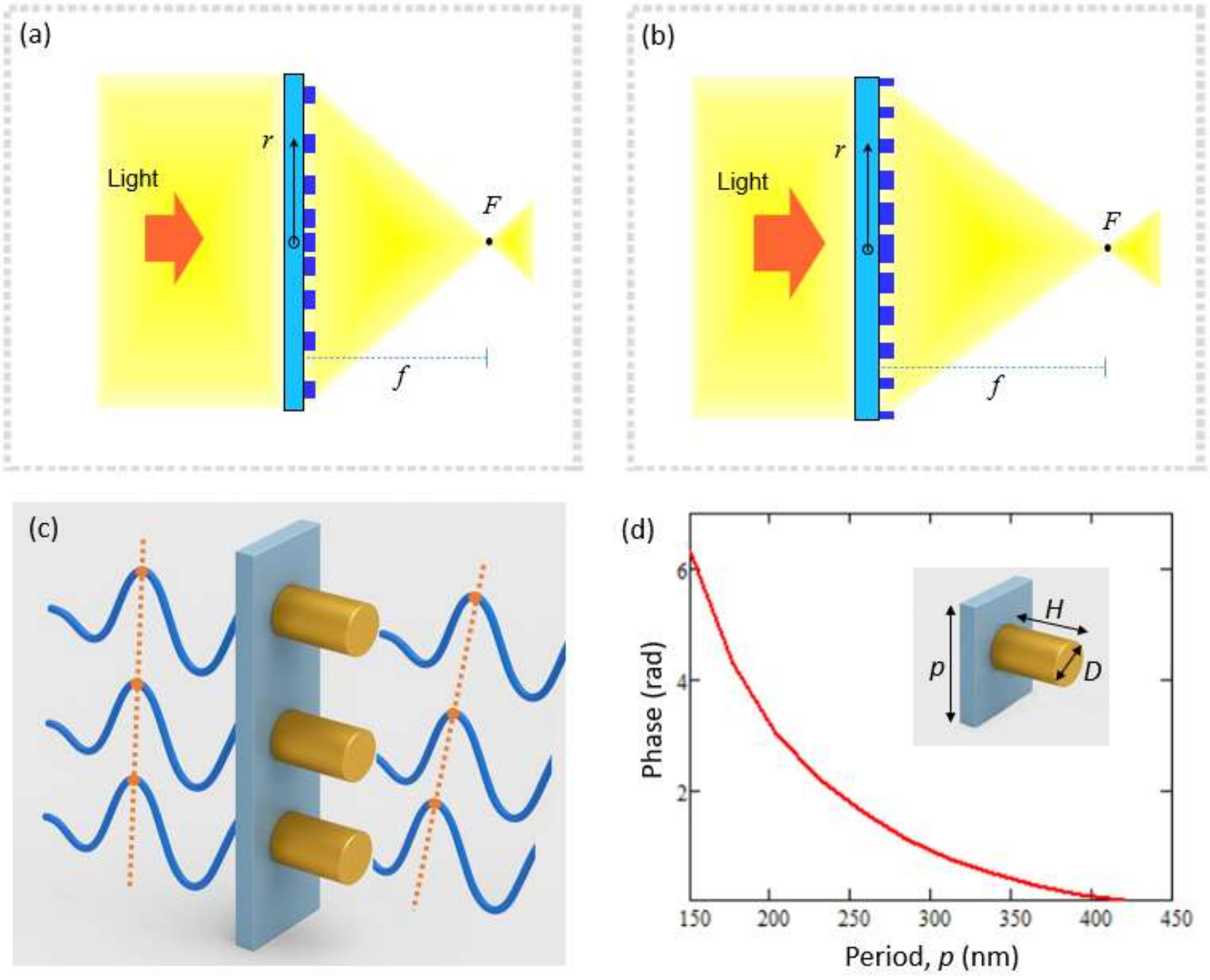

Fig. 1. Concept and working principle of the aperiodic metalens paradigm. (a) Schematic of the proposed aperiodic metalens, which achieves focusing using structurally identical nanocylinders with engineered variable periodicity. (b) A conventional metalens, which relies on varying nanocylinder diameters at a fixed period. (c) Illustration of the phase control mechanism: modulating the local period between identical nanocylinders directly controls the phase delay of the transmitted light. (d) Simulated phase shift as a function of the unit cell period ($p$), calculated via FDTD method, demonstrating full $2\pi$ phase coverage. The inset defines the unit cell parameters: nanocylinder diameter $D$, height $H$, and period $p$.

While the concept of modulating periodicity has been explored, prior work differs fundamentally from our single-parameter approach. The foundational idea was theoretically proposed by Gonidec for metallic nanobars in the infrared [11]; however, this approach was constrained by a phase range of less than $\pi$ and was not demonstrated in a complete metalens design. More recent efforts have employed hybrid strategies that combine a primary geometric variation—such as nanocylinder size or nanofin rotation—with a secondary modulation of the lattice periodicity, often guided by numerical optimization [12-16]. In these multi-parameter methods, periodicity is typically used as a corrective tool to optimize an existing size-variant (or orientation-variant) design, rather than as the fundamental phase-encoding mechanism. Rather than increasing the number of geometric degrees of freedom available to each meta-atom, here we explore the opposite limit: an invariant meta-atom whose optical phase is controlled exclusively through its spatial arrangement. We establish a new, deterministic metasurface paradigm that isolates the local period as the sole degree of freedom, thereby avoiding the need for complex geometry libraries. This fundamentally simple aperiodic architecture provides a robust, analytically tractable baseline for future state-of-the-art designs that may incorporate inverse optimization or free-form meta-atoms.

Aperiodic architectures exhibit the interesting feature of passive chromatic aberration suppression. The recent work by Palatnick et al. explored aperiodicity in the near-infrared using period and nanocylinder radius as free parameters within a numerical optimization—and reported that their results suggested an inherent association between aperiodicity and achromatic behavior [12]. Our approach explores this fundamental feature further by achieving visible-spectrum focusing using a single invariant building block—structurally identical nanocylinders—with local periodicity as the sole degree of freedom, demonstrating intrinsic near-achromaticity arising purely from the linear effective-index scaling of the architecture. This establishes a deterministic and analytically tractable metasurface paradigm that decouples phase from geometry without relying on complex geometry libraries or iterative optimization, and provides a reproducible physical baseline for future metasurface designs.

## 2. Working principle

### 2.1 Principle of phase modulation

The physical basis for most dielectric metalenses is the propagation phase mechanism, where the phase shift $\Phi$ is controlled by modulating the effective refractive index, $n_{\mathrm{eff}}$, of the unit cell. Conventionally, this is achieved by varying the meta-atom's diameter, $D$, while keeping the lattice period, $p$, constant, making the effective index a function of diameter, $n_{\mathrm{eff}}(D)$, which is characteristically lower than the refractive index of the nanorod material $n$, $n_{\mathrm{eff}}(D)<n$. Consequently, the phase shift is typically expressed as $\Phi(D) \approx (2\pi/\lambda)\, n_{eff}(D)\, H$, where $\lambda$ is the incident wavelength, and $H$ is the nanorod height [10,17]. Our work inverts this design philosophy. We investigate the complementary scenario: maintaining a fixed meta-atom geometry ($D$ and $H$ are constant) and instead engineering the phase response by exclusively varying unit cell or the local period, $p$.

The underlying principle is that the $n_{\mathrm{eff}}$ of the unit cell is not determined by the nanostructure alone, but by a hybridization of its optical properties with the surrounding medium. Modulating the period, $p$, directly alters the volume fraction of the high-index dielectric material within the unit cell. This, in turn, modifies the near-field coupling between adjacent nanostructures and manifests as a period-dependent effective refractive index, $n_{\mathrm{eff}}(p)$, as illustrated in Fig. 1(c). Consequently, the collective optical response depends on the periodicity: as the inter-element spacing increases, the transmitted wavefront delay decreases, and vice versa. This phase shift behavior can be expressed as:

$$\Phi(p) \approx \frac{2\pi}{\lambda} n_{eff}(p)\, H \qquad (1)$$

where $\lambda$ is the incident wavelength, and $H$ is the nanorod height. To validate this principle, we performed finite-difference time-domain (FDTD) simulations on a representative unit cell to calculate $\Phi(p)$. The results, exemplified in Fig. 1(d), confirm that a complete $2\pi$ phase shift can be achieved solely by varying the period $p$ within a practical range. This full phase coverage is the fundamental prerequisite for designing arbitrary metasurface based optics, including metalenses, and confirms the viability of controlling phase exclusively via periodicity.

### 2.2 Achromaticity principle

Chromatic aberration manifests as a frequency-dependent variation of the focal length. In metalenses, this effect arises from the combined influence of material dispersion, resonant scattering, and the intrinsically diffractive nature of metasurfaces [18–20], in contrast to refractive lenses where material dispersion is the dominant factor. For propagation-phase metalenses, dispersion plays a central role and is commonly analyzed by expanding the required phase profile in a Taylor series around the design frequency. In this framework, the zeroth-order term corresponds to the phase at the design frequency, while the first-order term represents the group delay. The group delay is defined as the derivative of the phase with respect to frequency, $\partial\Phi/\partial\omega$, evaluated at the design frequency $\omega_d$.

Within the propagation-phase regime, the phase profile can be approximated as $\Phi(r,\omega)\approx(\omega/c)n_{eff}(r,\omega)H$. Here $\omega$ is the angular frequency, c is the speed of light, $n_{eff}$ is the effective refractive index of the unit cell, $H$ is the pillar height, and $r$ denotes the radial coordinate from the metalens center [10,21]. Accordingly, the group delay is a function of the derivative of the effective refractive index with respect to frequency, $\partial n_{eff}(r,\omega)/\partial\omega$, evaluated at $\omega_d$.

Conventional achromatic metalens designs primarily focus on engineering this group delay and its higher-order dispersion. However, the focal position is not determined by the absolute phase profile, but by the radial phase gradient $\partial\Phi/\partial r$, which defines the local transverse wavevector ($k_r$) responsible for focusing. Consequently, chromatic aberration should be analyzed at the level of the spatial phase gradient rather than the phase itself.

In Section S1 (Supplemental Document), we present a theoretical analysis of focal achromaticity and derive a condition required for achromatic focusing:

$$\frac{\partial}{\partial r}\Phi(r,\omega) = \frac{\omega}{\omega_d}\frac{\partial}{\partial r}\Phi(r,\omega_d) \tag{2}$$

This condition agrees with the premise that the focal position is determined by the radial phase gradient rather than the absolute phase profile. The achromaticity condition implies that the radial phase gradient must scale linearly with frequency $\omega$ with respect to the design phase gradient. One must note that this condition is not only function of $\omega$ but also of $r$. Deviations from this achromaticity condition result in a frequency-dependent focal shift whose magnitude scales with the degree of deviation.

For propagation-phase metalenses, this condition can be reformulated in terms of the effective refractive index using effective medium theory [22-25]. Within this framework, the desired radial phase profile is implemented through a spatial modulation of the unit-cell fill factor $\gamma$, which is made to vary as a function of the radial coordinate $r$. Here, the fill factor is defined as the ratio between the nanopillar volume and the unit-cell volume, $\gamma=V_{pillar}/V_{cell}$. As shown in Section S1, achromatic focusing is achieved when the derivative of the effective refractive index with respect to the fill factor $\gamma$ remains invariant with frequency $\omega$ over the operational bandwidth, i.e.,

$$\frac{\partial}{\partial\gamma}\, n_{eff}(\gamma,\omega) = \frac{\partial}{\partial\gamma}\, n_{eff}(\gamma,\omega_d) \tag{3}$$

This condition reveals the fundamental dispersive origin of chromatic behavior in propagation-phase metalenses: it is not the geometry per se, but how the effective index responds to the fill factor. More precisely, the suppression of chromatic aberration is determined by the frequency dispersion of the effective-index gradient, rather than by the dispersion of the absolute effective index. Consequently, the proposed geometry-invariant architecture inherently satisfies this condition to a much greater degree than conventional size-variant architectures, simply because the dispersive material volume comprising each nano-element remains constant across the entire metasurface. From a waveguide perspective, varying the diameter of the nanoelements in conventional periodic metalenses modifies mode confinement and the associated effective index nonlinearly, making the slope $\partial n_{eff}/\partial \gamma$ more dependent on both the local fill factor and frequency than in the aperiodic architecture. Thus, the observed chromatic behavior arises from the intrinsic dispersion of the phase-encoding mechanism rather than from explicit dispersion engineering.

This physical insight is rigorously validated via FDTD simulations in Section S1, which demonstrate that the aperiodic metalens approximates the ideal achromatic focusing condition more closely than its periodic counterpart. Importantly, these results provide a fundamental physical basis that is in agreement with recent reports on the achromatic potential of combining aperiodicity with classical periodic metasurfaces [12]. Specifically, we show that by maintaining a constant nanorod diameter, the effective refractive index scales linearly with the local fill factor ($\gamma$). This intrinsic linearity ensures that the derivative $\partial n_{eff}/\partial \gamma$ is independent of the fill factor and exhibits low frequency dependence, in contrast to classical periodic metalenses, where this derivative strongly depends on both parameters. Ultimately, this invariant derivative constitutes the physical mechanism enabling the passive chromatic aberration suppression characteristic of our design.

The intrinsic near-achromatic performance of the aperiodic architecture stems from the robustness of the truncated waveguide mechanism when utilizing identical nanocylinders. By fixing the nanopillar diameter and modulating only the lattice spacing, the intrinsic modal dispersion of the waveguides remains approximately invariant across the metalens. This contrasts sharply with periodic designs, where varying the diameter alters the optical confinement factor and the waveguide dispersion. In our approach, the geometric invariance leads to a linear dependence of the effective refractive index on the fill factor, which implies that the phase gradient scales nearly proportionally with frequency. In the context of dispersion engineering, this behavior naturally linearizes the spatial group delay profile while suppressing higher-order group delay dispersion terms. Consequently, the required focal condition is, in principle, satisfied over a broader bandwidth without the phase errors typically introduced by the nonlinear dispersive response of variable-diameter waveguides, resulting in an intrinsically higher achromatic focusing behavior.

## 3. Metalens design: deterministic coordinate generation

An important advantage of the proposed design paradigm is its deterministic and analytically tractable nature. Because the meta-atom geometry is fixed, the design problem reduces to determining the spatial coordinates of identical nanocylinders according to the prescribed phase profile. As shown in this section, these coordinates can be generated deterministically from the local periodicity, without numerical optimization of the metalens layout. This approach therefore provides a direct and reproducible route from the desired phase profile to the physical arrangement of the meta-atoms.

Generally, to focus normally incident light, a metalens must impart a hyperbolic phase profile, $\Phi(r)$, given by:

$$\Phi(r) = -\frac{2\pi}{\lambda}\left(\sqrt{r^2 + f^2} - f\right) \tag{4}$$

where $\lambda$ is the design wavelength of light in free space, $f$ is the focal length, and $r$ is the radial distance from the metalens center. Eq. (4) is the standard phase profile for aberration-free on-axis focusing; however, other phase profiles may be required for cascaded metalenses or specific applications [26]. Once the phase profile is defined, the metalens is implemented by determining the spatial arrangement of the meta-atoms. Conventional metalenses are typically designed by employing a periodic array of nanocylinders and varying their diameters, $D$, to match the phase distribution in Eq. (4). However, while conventional metalenses realize the required phase profile by assigning a specific nanorod diameter $D(r)$ to a radial position $r$, our approach assigns a local period $p(r)$ to $r$. Consequently, the aperiodic metalens is constructed by arranging identical nanocylinders with a spatially varying period $p$. Since all meta-atoms are structurally identical, the design process simplifies to determining the precise spatial coordinates $(x_i, y_j)$ of each nanocylinder.

The first step is to establish the relationship between the local period $p$ and the imparted phase $\Phi$, i.e., the function $p(\Phi)$. This requires identifying a unit cell that achieves full $2\pi$ phase coverage with high transmission solely by varying its period. The range of periods ($p_{min}$=150 nm to $p_{max}$=420 nm) was carefully bounded to ensure optimal optical performance. The minimum period ensures sufficient separation between the 120 nm nanopillars to prevent severe near-field coupling and maintain the validity of the local phase approximation. Conversely, the maximum period is strictly limited to satisfy the subwavelength Nyquist criteria ($p<\lambda$/NA) required to suppress the emergence of efficiency-degrading higher-order diffraction modes.

For simplicity, we utilized a square unit cell containing a single nanocylinder (composed of either Si or $TiO_2$) on a fused silica ($SiO_2$) substrate (Fig. 2(a)) for operation in the visible spectrum. The phase shift and light transmission were calculated using the FDTD method (Figs. 2(b)-(e)). In these simulations, a plane wave linearly polarized along the $x$-axis was incident from the substrate side, with periodic boundary conditions applied in the transverse ($xy$) plane and perfectly matched layers along the propagation ($z$) axis. Through a comprehensive parametric sweep, calculating phase and transmission by varying the height and period for different nanocylinder diameters, we identified an optimal geometry with a diameter $D$ = 120 nm and a height $H$ (800 nm for Si rods, and 1200 nm for $TiO_2$ rods), which yielded complete phase coverage and high transmission at the design wavelength. In other words, the meta-atom design process explored different $H$ and $D$ values via simulations to generate phase and transmission maps as functions of $H$ and $p$, leading to the identification of the suitable $H$ and $D$ values. A similar process was performed for conventional (periodic) meta-atoms, where the suitable height $H$ was 350 nm for Si rods (650 nm for $TiO_2$ rods), and the fixed period $p$ was 300 nm for both Si and $TiO_2$ configurations. Figures 2(b)-(e) were obtained using the refractive index and extinction coefficient at the design wavelength (660 nm). However, in Section 4, the metalens chromatic performance is rigorously analyzed using the full spectral parameters. In other words, all wavelength-dependent FDTD simulations employed the full complex refractive index of silicon and $TiO_2$, $n(\lambda)+ik(\lambda)$, whose dispersion curves are provided in Fig. 2(f).

The resulting phase and transmission curves for our selected meta-atom as a function of period are shown in Figures 2(b)-(c). These results confirm that a full $2\pi$ phase coverage is achieved as the period varies from 150 nm to 420 nm, while maintaining high average transmission. Crucially, the phase response is monotonic and free of sharp resonances (which are typically prevalent in conventional size-variant meta-atoms, as seen in Figs. 2(d)-(e)). This smooth, deterministic relationship is highly advantageous for metalens design, as it enables a precise mapping from the target phase to the required local periodicity, mitigates performance degradation from resonant effects, and enhances the accuracy of advanced modeling techniques such as metasurface ray-tracing. In other words, the phase response follows a smooth, continuous curve devoid of resonances, exhibiting an inverse proportionality to the lattice period $p$. This predictable behavior potentially enhances the design accuracy for advanced optical systems, such as cascaded metalenses [26-28].

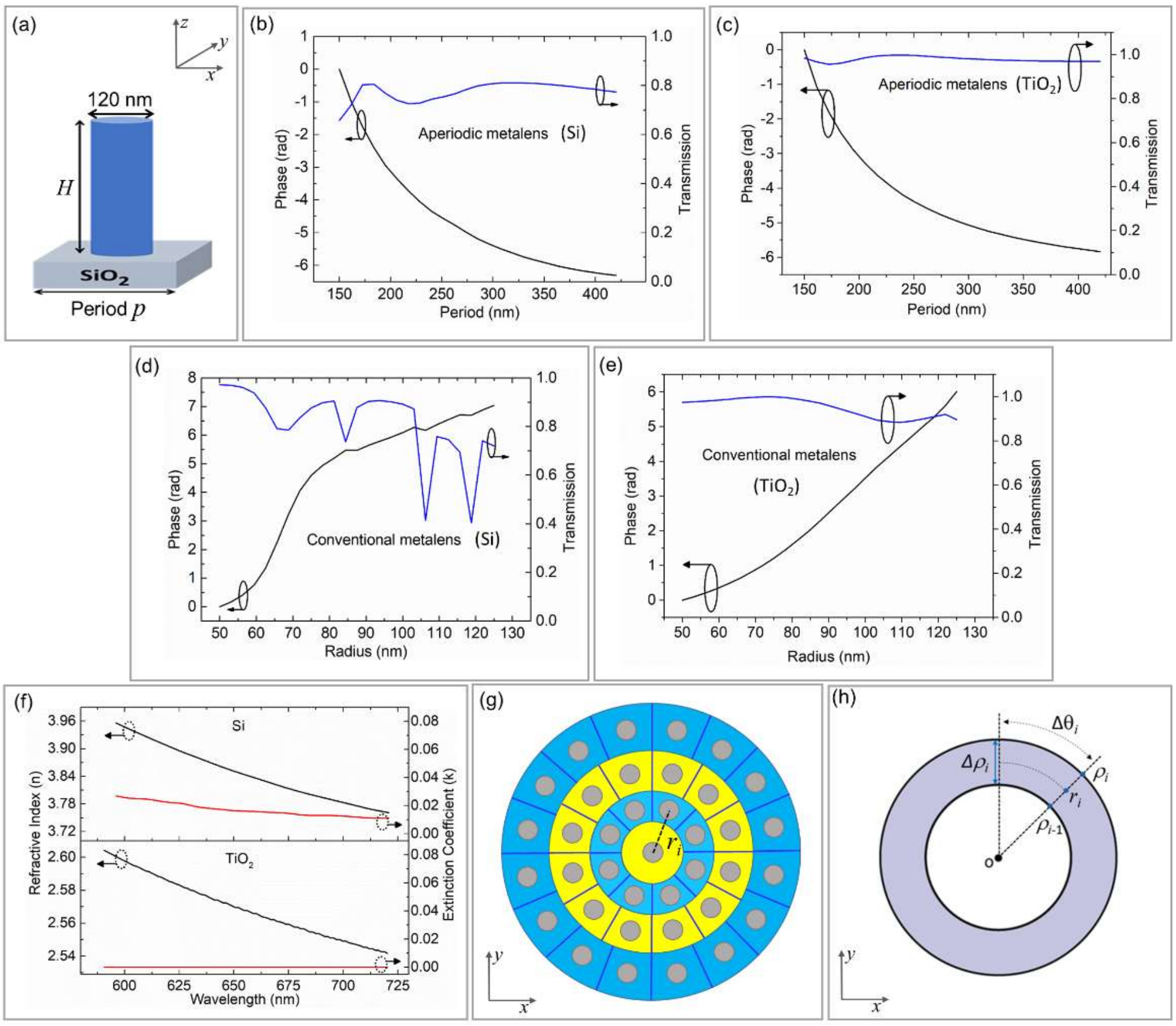

Fig. 2. (a) Schematic representation of the unit cell with period $p$ and its cylindrical nanopillar. We consider pillars composed of either silicon (Si) or titanium dioxide ($TiO_2$). The base substrate is square, with the side length equal to the period $p$. (b) and (c) show the plot of the phase $\Phi(p)$ and transmission $T(p)$ for Si and $TiO_2$ aperiodic meta-atoms, respectively. This is the phase shift as a function of the unit cell period (black line) and light transmission (blue line) as a function of the unit cell period. These plots are computed at the design wavelength (660 nm), using as refractive index and extinction parameters $n$=3.835 and $k$=0.016 for Si, and $n$=2.565 and $k$=0 for $TiO_2$. The corresponding phase and transmission plots as a function of nanorod radius for conventional meta-atoms are shown in (d) for Si and (e) for $TiO_2$. (f) Spectral dispersion curves (refractive index and extinction coefficient) for Si and $TiO_2$. These spectral values were used for all chromatic focusing performance analyses in Section 4. (g) Metalens design strategy inspired by mosaic patterning. The metalens surface is divided into concentric rings (blue and yellow), each further partitioned into segments that serve as unit cells (defining the unit-cell grid), with each cell containing its corresponding nanorod (gray) at polar coordinates $(r_i,\theta_{ij})$. (h) Schematic of a single concentric ring, illustrating its partition into unit cells with angular and radial dimensions $\Delta\theta_i$ and $\Delta\rho_i$, respectively. This segmentation is used to determine the precise coordinates $(r_i,\theta_{ij})$ of each nanorod.

Once the $\Phi(p)$ curve is obtained, the required aperiodicity $p(r)$, with $r$ denoting the radial distance from the metalens center, can be determined. To implement the hyperbolic phase profile $\Phi(r)$ given by Eq. (4), a specific spatial period distribution $p(r)$ must be prescribed. The first step is to extract the function $p(\Phi)$, i.e., the period as a function of phase, which we obtain by numerically fitting the data in Fig. 2(b) (or Fig. 2(c)). Once the $p(\Phi)$ is known, the spatial period profile follows directly from the analytical phase function $\Phi(r)$, yielding i.e. $p(r)= p[\Phi(r)]$. This straightforward procedure leads to Eq. S13 for Si meta-atoms or the equivalent for $TiO_2$ meta-atoms (Supplemental Document).

Knowing $p(r)$ enables the next step: filling the surface with unit cells of appropriately varying size. This process is analogous to assembling a mosaic composed of distinct pieces (tesserae) that collectively form a complete pattern. The challenge lies in determining the shape, size, and arrangement of these unit cells to realize the desired metalens design. This mosaic analogy also highlights the large design space available, since many distinct tessellations are possible. In our case, the unit cells used in simulations have square shape with varying lateral dimensions (Fig. 2a), which makes a direct tiling of the metasurface appear nontrivial. However, this difficulty can be alleviated by adopting the hypothesis that the phase delay remains approximately invariant under slight geometric distortions of a unit cell, provided that its cross-sectional area is preserved [10,29,30].

Under this assumption, we employ a simple and robust design strategy: the metalens surface is divided into concentric rings, each further subdivided into segments that act as individual unit cells (Fig. 2(g)). As the radial distance $r$ increases, these segments naturally approximate rectangular or square shapes. To keep them as square as possible, we impose a "squareness" condition derived in Section S3 (Supplemental Document). This "squareness" condition leads to the required radial coordinates $r_i$ for each nanorod, which are determined by solving the following nonlinear equation for each ring:

$$\rho_i - \rho_{i-1} - p(r_i) = 0 \quad , \tag{5}$$

where $\rho_i$ and $\rho_{i-1}$ are the outer and inner radii defining the $i$-th ring (see Fig. 2(h)), $i$=1…$N$. The period profile $p(r)$ is given by Eq. (S13), and the radial position of nanorods within each ring is simply

$$r_i = (\rho_i + \rho_{i-1})/2 \quad , \tag{6}$$

Each ring is then divided into equal angular segments of size $\Delta\theta_i = p(r_i)/r_i$. Accordingly, the coordinates of each nanorod are $(r_i, \theta_{ij})$, where $\theta_{ij}=j\Delta\theta_i$ or equivalently:

$$\theta_{ij} = j\frac{p(r_i)}{r_i} \quad , \tag{7}$$

with $j$=0…$M_i$, where $j$ indicates the $j$-th ring segment in each ring, and $M_i$ is defined in Section S3 (Supplemental Document).

This design procedure requires solving Eq. (5) $N$ times to determine the radial coordinates of the $N$ rings of nanorods. Although this equation is evaluated numerically, the procedure is fully deterministic, with no iterative optimization or parameter search involved. Figure 3 shows four examples of spatial distributions generated by this method, illustrating the implementation of the proposed metalens architecture: an aperiodic arrangement of identical nanocylinders whose positions are entirely determined by Eqs. (5)–(7). In other words, the metalens layout is obtained by repeatedly solving Eq. (5) and by computing $(r_i, \theta_{ij})$ from Eqs. (6) and (7).

The design method therefore constitutes an analytical framework rather than a numerical optimization procedure. Because the meta-atom geometry is fixed, the complete layout is determined by a straightforward set of equations that directly maps the prescribed phase profile onto the spatial coordinates of the identical nanocylinders. Coordinates and coordinate-generation scripts are provided in Sections S4 and S5 of the Supplemental Document to facilitate full reproducibility. The resulting layouts form the basis of the metalenses whose optical performance is analyzed in the following section.

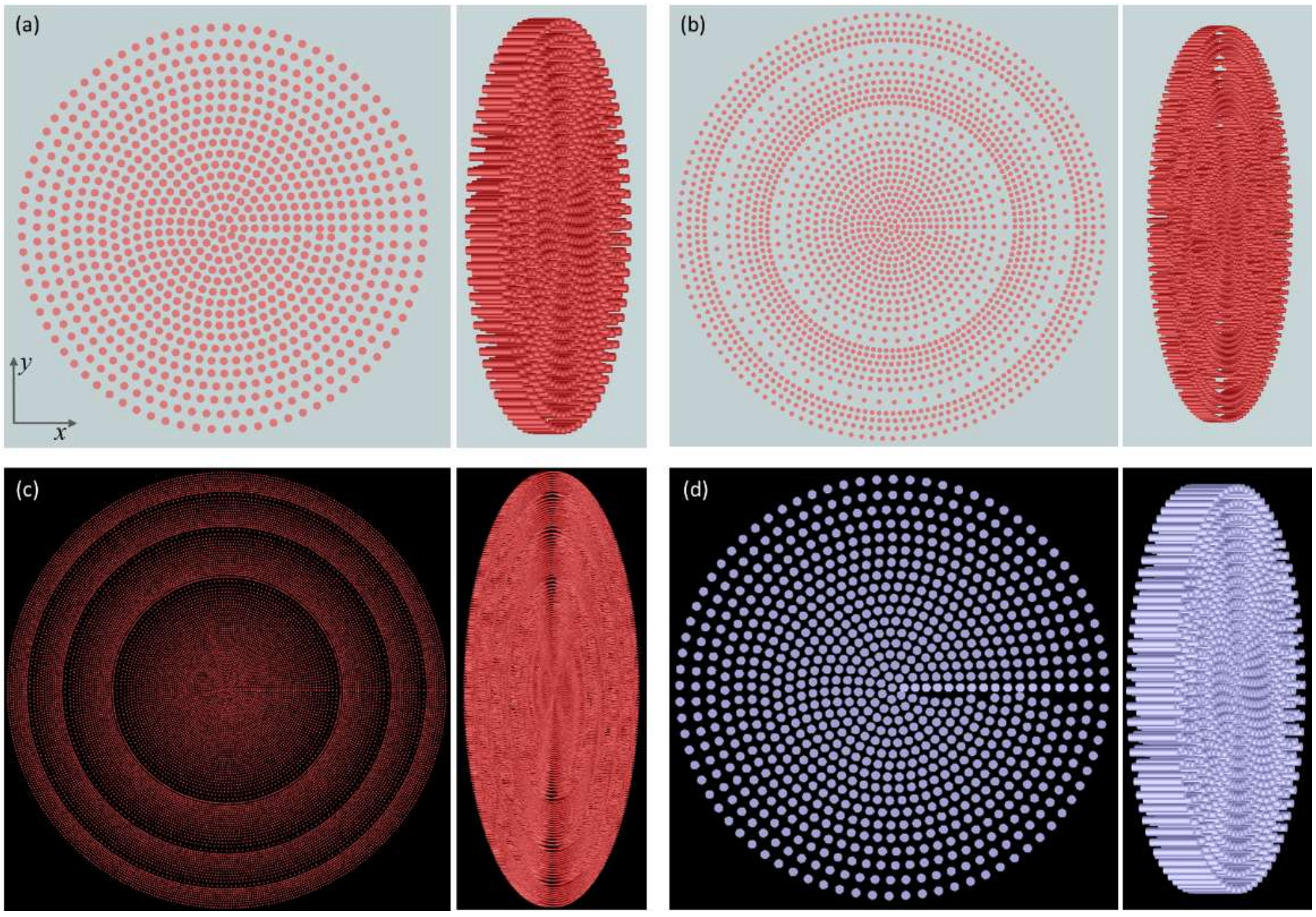

Fig. 3. Layouts of the proposed aperiodic metalenses composed of identical nanocylinders. Top and side views illustrate the structure and spatially varying lattice periodicity. (a) Moderate numerical aperture (NA) design ($f$=9.9 µm, diameter $D_m$=5.9 µm, NA=0.29) consisting of $N$=17 rings. (b) High-NA design ($f$=5.94 µm, diameter $D_m$=10.1 µm, NA=0.845) consisting of $N$=26 rings. (c) Layout of the large-diameter design ($f$=60 µm, diameter $D_m$=35 µm, NA=0.3) consisting of $N$=90 rings. (d) Layout of the $TiO_2$ metalens design ($f$=9.9 µm, diameter $D_m$=5.8 µm, NA=0.29) consisting of $N$=17 rings.

## 4. Numerical validation of the aperiodic metalens

This section evaluates the proposed design paradigm and its passive suppression of chromatic aberration through full-wave FDTD simulations. The numerical study serves as a proof of concept, demonstrating that the proposed architecture can achieve focusing while reducing chromatic focal shift without explicit dispersion engineering of the meta-atoms. We compare the optical response of the aperiodic and conventional periodic architectures under identical target specifications: design wavelength $\lambda_d$=660nm, focal length $f$, aperture diameter $D_m$, and constituent materials. Four representative cases are considered, covering moderate and high numerical apertures (NA= 0.4 and 0.8), large scalable apertures, and two dielectric platforms (Si and $TiO_2$). The first three cases use Si nanocylinders, while the fourth uses $TiO_2$; all metalenses are placed on a $SiO_2$ substrate. The proposed aperiodic metalenses consisted of identical nanorods with sizes: $D$=120 nm, and $H$=800 nm (for Si rods) or $H$=1200 nm (for $TiO_2$ rods). The spatial coordinates of the nanorods for the first two cases are provided in Section S4, and the complete Python/MATLAB code for their calculation is provided in Section S5 of the Supplemental Document. For comparison, the conventional size-variant metalenses were designed using nanorods of variable diameter ($D$=100–250 nm) and a different height ($H$=350 nm for Si rods, and $H$=650 nm for $TiO_2$ rods) arranged on a fixed 300 nm period lattice. In the simulations, each metalens was positioned at the $z$=0 plane and illuminated by a normally incident plane wave propagating along the $+z$ direction. The transmitted field was monitored to analyze the focusing behavior and its chromatic response across the visible spectrum, using the full wavelength-dependent dispersive optical constants of each material (Fig. 2(f)).

### *4.1 Metalens with moderate numerical aperture*

Here we evaluate the optical performance of an aperiodic metalens with moderate NA. For the analysis presented in this section, both the aperiodic and conventional periodic metalenses were designed with identical target optical parameters: $f$=9.9µm, and an aperture diameter of 5.9 µm. The proposed aperiodic metalens consists of structurally identical nanorods arranged in 17 concentric rings. The focal lengths resulting from simulations were 6.95µm for the conventional metalens (NA≈0.42), and 7.07µm for the aperiodic metalens (NA≈0.42).

Figure 4 presents a comprehensive analysis of the monochromatic focusing performance at the design wavelength (660 nm). Figure 4(a) shows the 1D intensity profiles for the aperiodic metalens, depicting the transverse ($x$-axis) and longitudinal ($z$-axis) distributions through the focal spot, with corresponding 2D intensity maps as insets. Figures 4(b) and 4(c) provide the 2D intensity distributions in the focal ($xy$) plane (top) and cross-sectional ($xz$) plane (bottom) for the aperiodic and conventional metalenses, respectively.

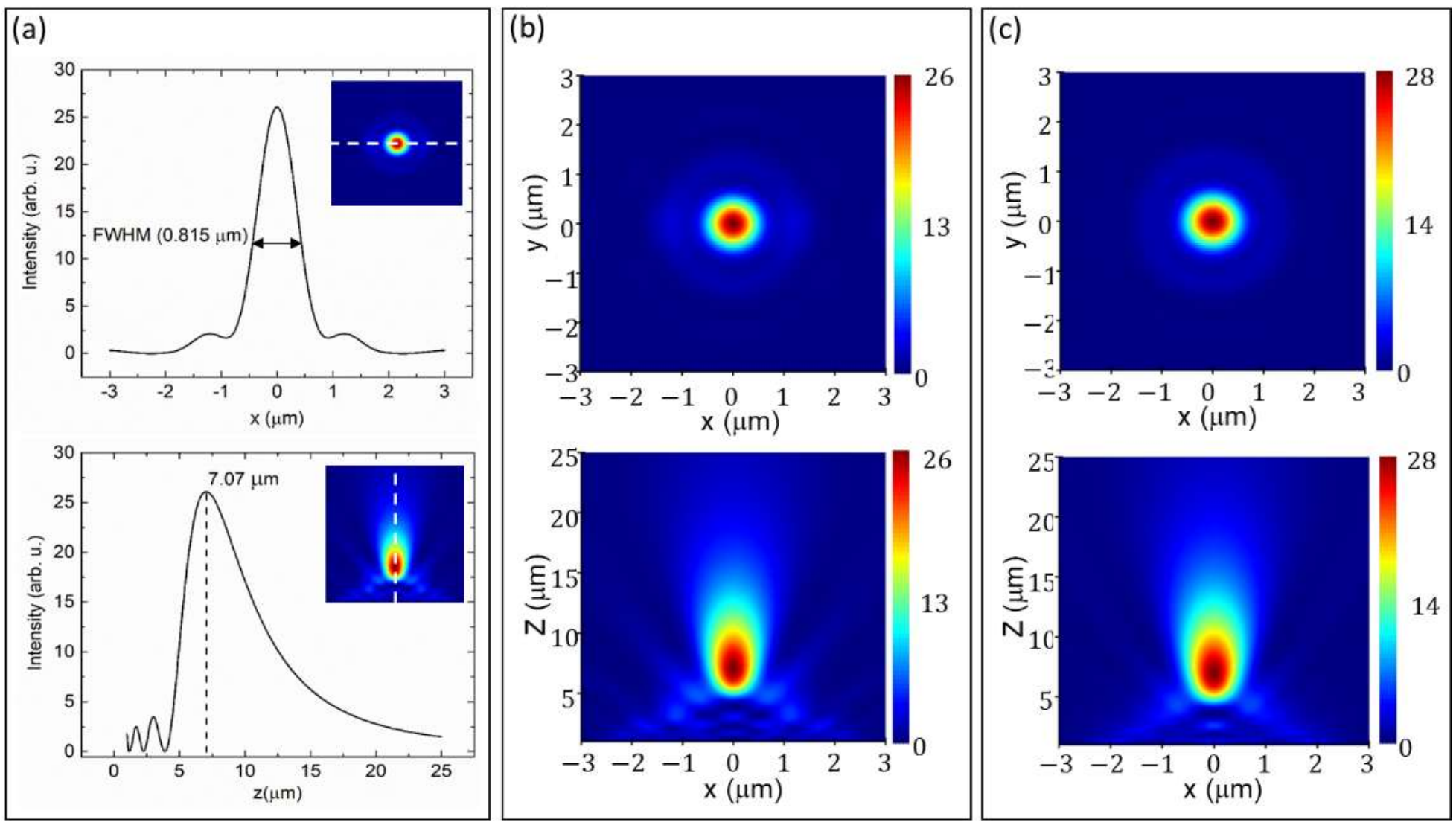

Fig. 4. Comparative analysis of monochromatic focusing performance. FDTD simulations comparing the proposed aperiodic metalens and a conventional size-variant metalens at the design wavelength ($\lambda_d$ = 660 nm). (a) Transverse (top) and longitudinal (bottom) 1D intensity profiles through the focal spot of the aperiodic metalens, with 2D intensity maps shown as insets. (b, c) Corresponding 2D intensity distributions in the focal ($xy$) and cross-sectional ($xz$) planes for the (b) aperiodic and (c) conventional metalenses, respectively. Both metalenses were designed with equivalent optical parameters (identical focal length and aperture). The simulated focal lengths were 6.95µm for the conventional metalens (NA≈0.42), and 7.07µm for the aperiodic metalens (NA≈0.42).

While the 2D focusing profiles in Fig. 4(b) and 4(c) appear qualitatively similar, a quantitative analysis of the focal spots reveals a superior performance for the aperiodic design. A basic quantitative metric for evaluating metalens focusing is the full width at half maximum (FWHM) [31]. Our aperiodic metalens achieves a focal spot with a FWHM of 0.815 µm, which is only 1.4% larger than its theoretical diffraction limit ($\lambda/2NA$) of 0.804 µm. In contrast, the conventional metalens produces a larger focal spot with an FWHM of 0.871 µm, approximately 12% above its diffraction limit of 0.777 µm. This result provides clear evidence that the aperiodic design paradigm not only simplifies the structure but also enhances the focusing resolution, achieving a tighter, more diffraction-limited spot.

To characterize the chromatic performance, additional FDTD simulations were carried out as a function of wavelength. We computed the focal length and focusing efficiency over a 120 nm bandwidth (600–720 nm) centered on the design wavelength. The results, presented in Figure 5, reveal a significant reduction in chromatic focal dispersion for the aperiodic design. The root mean squared error (RMSE) of the focal length is reduced from 1.4 µm for the conventional metalens to just 0.8 µm for the aperiodic metalens, a reduction of nearly 42%. Furthermore, the aperiodic metalens exhibits a markedly smoother spectral dependence of the focusing efficiency (Fig. 5(c)). This behavior confirms that the mechanisms described in Section 2.2 results from employing identical nanocylinders across the surface. As a consequence, the spectral response becomes more uniform and predictable. The slightly lower overall efficiency of the aperiodic design ($H$ = 800 nm) compared to the conventional one ($H$ = 350 nm) is ascribed to the greater material volume, which leads to increased absorption—particularly relevant when using a material with non-negligible extinction coefficient ($k$=0.016 at $\lambda_d$=660 nm). In Section S6 of the Supplementary Document, we show that, when both designs use nanocylinders of the same height, the aperiodic metalens achieves a higher average focusing efficiency and a 48% lower RMSE of the focal length. We also observe inverse and complementary spectral trends in the focusing properties of the two designs (Fig. S4).

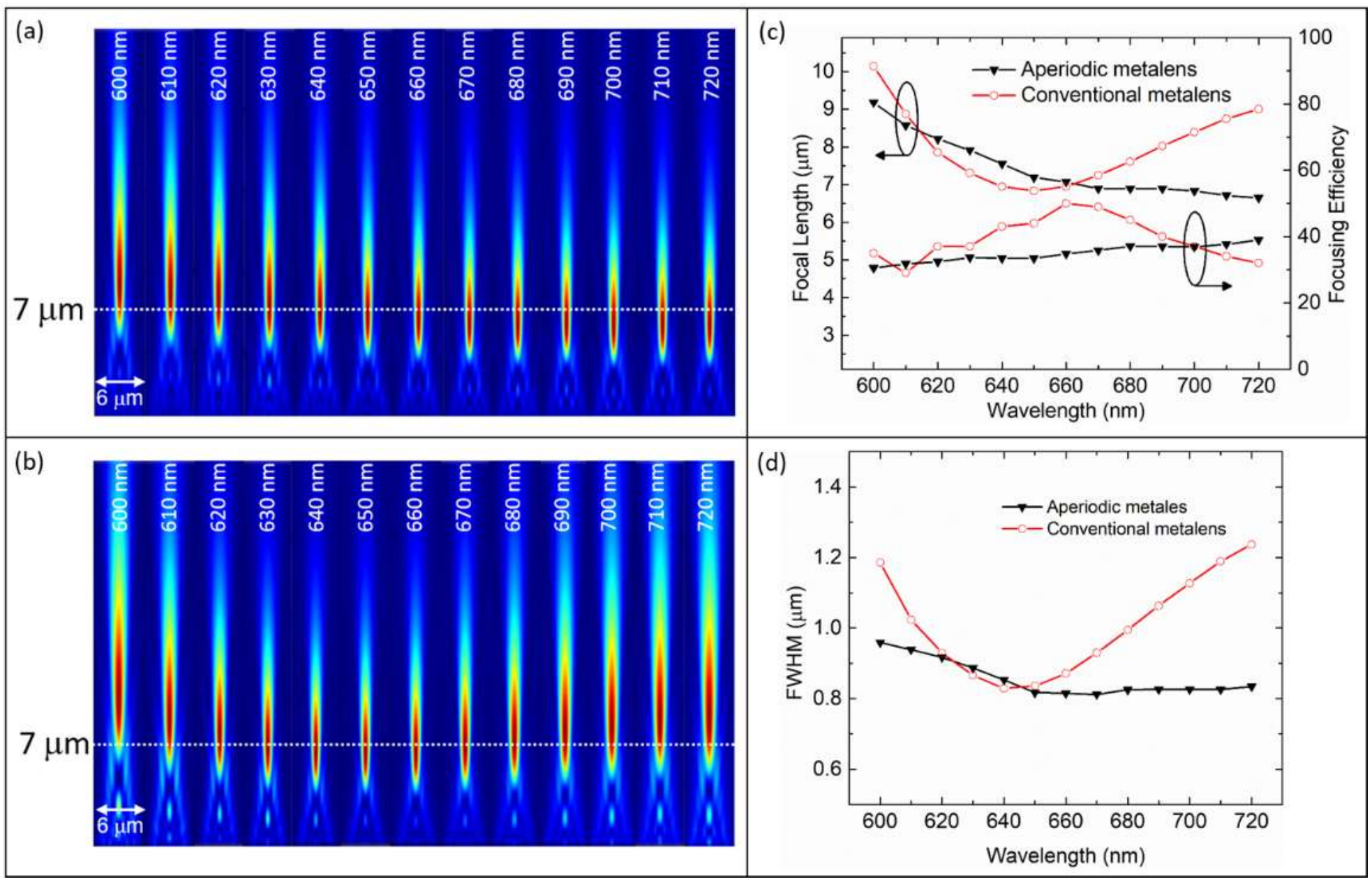

Fig. 5. Chromatic focusing performance analysis. Longitudinal intensity distributions ($xz$-plane) simulated across the 600–720 nm spectral range for (a) the aperiodic metalens and (b) the conventional size-variant design, illustrating the chromatic focal shift. (c) Quantitative comparison of focal length and focusing efficiency as a function of wavelength. (d) Full Width at Half Maximum (FWHM) of the focal spot versus wavelength. The metalenses analyzed here correspond to the monochromatic designs presented in Fig. 4, which were designed for $\lambda_d$=660 nm and are evaluated here under broadband illumination.

Achromatic operation of a metalens can be achieved if the variation in the focal length ($\Delta f = f_{max} - f_{min}$) at both ends of the working bandwidth is smaller than the depth of focus (DoF) of the metalens [32], i.e., $\Delta f$ /DoF<1. The DoF can be defined as the longitudinal size where the electric field intensity along the optical axis (z-axis) is greater than 80% of the maximum focused intensity [33,34], or when the intensity remains above 50% of the maximum [35]. Under the $I(z) > 0.8 I_{max}$ criterion, the DoF for a diffraction limited metalens is $\lambda/NA^2$ [33]. In the case of the aperiodic metalens, the focal shift $\Delta f$=2.45 µm is smaller than that of the conventional metalens ($\Delta f$=3.1 µm), and both fall within the theoretical DoF ($\lambda/NA^2$) of 3.74 µm. The DoF naturally varies at each wavelength; the complete set of spectral DoF values obtained for this configuration, as well as for the subsequent metalens designs, is detailed in Section S7 of the Supplemental Document.

At first glance, Figs. 5(a) and (b) suggest an interesting property of the chromatic focusing behavior of the aperiodic metalens: it appears to maintain a sharper focal spot across the spectrum. This can be quantified by examining the FWHM of the focal spot. Figure 5(d) presents the FWHM values for both the aperiodic and the conventional metalenses over the same spectral range (600–720 nm). The aperiodic metalens consistently outperforms the conventional design, exhibiting a significantly smaller focal-spot width, with an average FWHM of 850 nm compared to 1000 nm for the traditional metalens.

The higher spectral stability of the aperiodic design is associated with its suppression of size-related resonance effects, which typically arise at nanorod dimensions near $\lambda/n$, where $n$ is the refractive index. In the aperiodic metalens analyzed here, all nanorods share the same diameter (120 nm), resulting in a single dominant resonance near $\lambda \approx 510$ nm for $n$=4.2485. This behavior is illustrated in Fig. 6, which shows the intensity profiles at the focal plane (*xz*-plane) across the entire visible spectrum (400–720 nm) for the same design wavelength $\lambda_d$=660 nm. Accordingly, the aperiodic metalens exhibits an atypical response near the resonance wavelength. A similar, though weaker, effect occurs near $\lambda \approx 510$nm in the conventional metalens, whose nanorod diameters vary radially from 100 to 250 nm, introducing multiple size-dependent resonances associated with the various rod dimensions distributed across the metalens aperture.

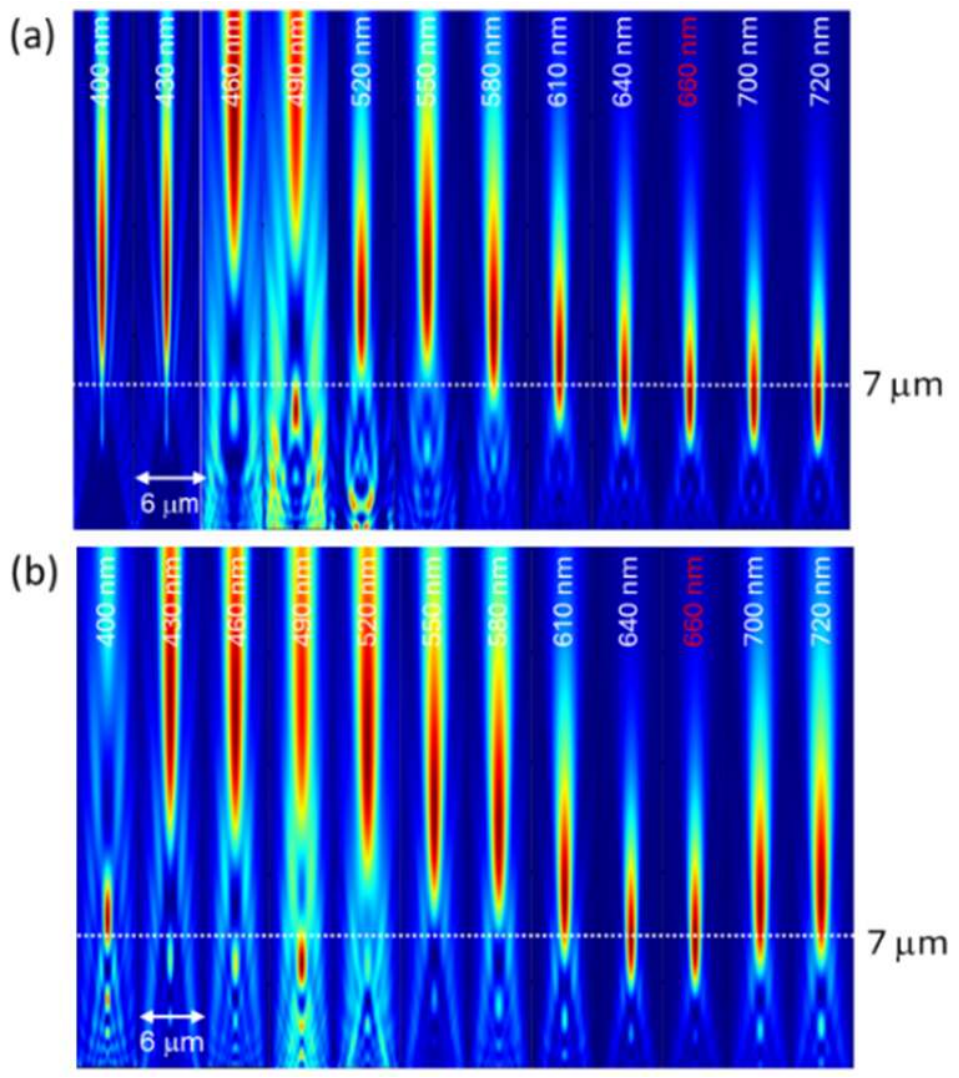

Fig. 6. Intensity profiles at the focal point through the *xz*-plane for (a) the aperiodic, and (b) a conventional metalens across the entire visible spectrum (400 – 720 nm). Both metalenses were designed for $\lambda_d$ = 660 nm.

The key performance metrics for both designs are summarized in Table S3 (Section S8 of the Supplemental Document) for direct comparison. The aperiodic architecture exhibits reduced chromatic focal shift, tighter average focal spots, and more stable efficiency across the operational bandwidth compared with the conventional geometry-variant design. Importantly, these results are obtained using a deterministic, unoptimized design based on a single nanocylinder geometry. This provides a simple baseline from which further performance improvements could be pursued through established optimization strategies [36], while preserving the proposed phase-encoding principle based on identical building blocks.

### *4.2 Metalens with high numerical aperture*

This section evaluates the optical performance of an aperiodic metalens designed with a high NA. To provide a fair comparison, both the aperiodic and conventional periodic metalenses were designed with identical target optical parameters: $f$=5.94μm, and an aperture diameter of 10.1 μm. The proposed aperiodic metalens consists of structurally identical nanorods arranged in 26 concentric rings. Simulations yielded an effective focal length of 6.35μm (NA=0.8) for the aperiodic design, and 6.41μm (NA=0.79) for the conventional counterpart at the design wavelength (660 nm).

Figure 7 presents their monochromatic focusing performance. Figure 7(a) illustrates the 1D intensity profiles for the aperiodic metalens, depicting the transverse ($x$-axis) and longitudinal ($z$-axis) distributions through the focal spot, with corresponding 2D intensity maps as insets. Figures 7(b) and 7(c) display the 2D intensity distributions for the aperiodic and conventional metalenses, respectively.

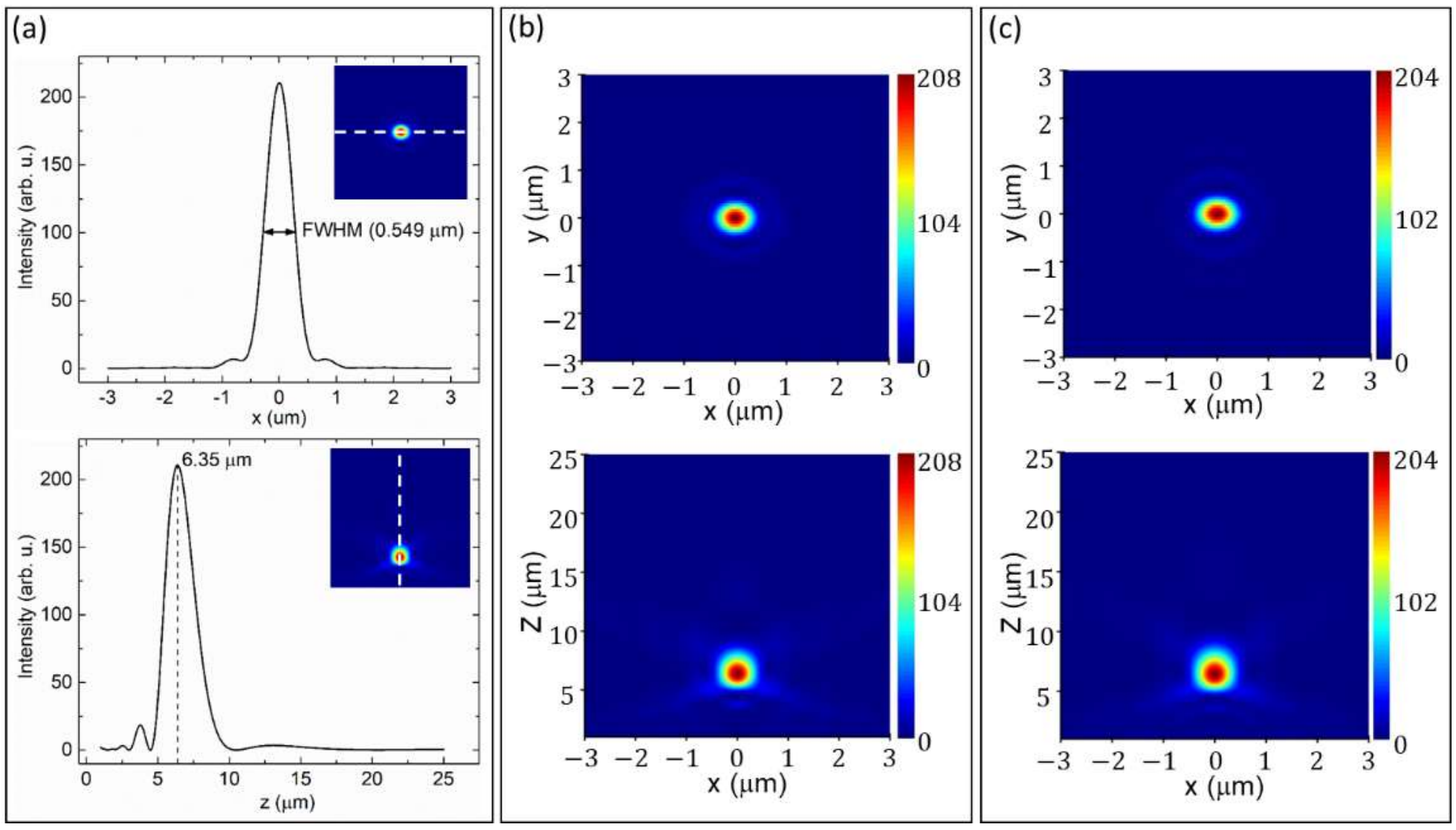

Fig. 7. Monochromatic focusing performance of high-NA metalenses at the design wavelength ($\lambda_d$=660 nm). (a) Transverse (top) and longitudinal (bottom) 1D intensity profiles through the focal spot of the aperiodic metalens, with 2D intensity maps shown as insets. (b, c) Corresponding 2D intensity distributions in the focal ($xy$) and cross-sectional ($xz$) planes for the (b) aperiodic and (c) conventional metalenses. Both initial designs share identical target optical parameters. The simulated focal lengths are 6.41μm for the conventional metalens (NA=0.79), and 6.35μm for the aperiodic metalens (NA=0.8).

Qualitatively, the focusing profiles in Fig. 7 appear similar; however, a closer inspection reveals that the aperiodic metalens produces a tighter focal spot. Specifically, the aperiodic design achieves a FWHM of 0.549 μm, which is 30% larger than its theoretical diffraction limit ($\lambda/2NA$) of 0.415 μm. In contrast, the conventional metalens produces a larger focal spot with a FWHM of 0.593 μm, deviating by 42% from its diffraction limit (0.419 μm). These results provide clear evidence that, even in the high-NA regime, the aperiodic design paradigm naturally enhances focusing resolution, approaching the diffraction limit more closely than the conventional approach.

The chromatic performance of both designs is analyzed in Fig. 8 over a 120 nm bandwidth (600–720 nm) centered on the design wavelength. Visually, Figs. 8(a) and 8(b) suggest similar chromatic focal dispersion behaviors for both architectures. Quantitatively, the longitudinal chromatic aberration is comparable, as shown in Fig. 8(c); the RMSE of the focal length variation is 0.5 μm for the conventional metalens and 0.56 μm for the aperiodic metalens. Regarding efficiency, the conventional design exhibits higher focusing efficiency at wavelengths below 690 nm, while the aperiodic metalens outperforms it at wavelengths above 690 nm.

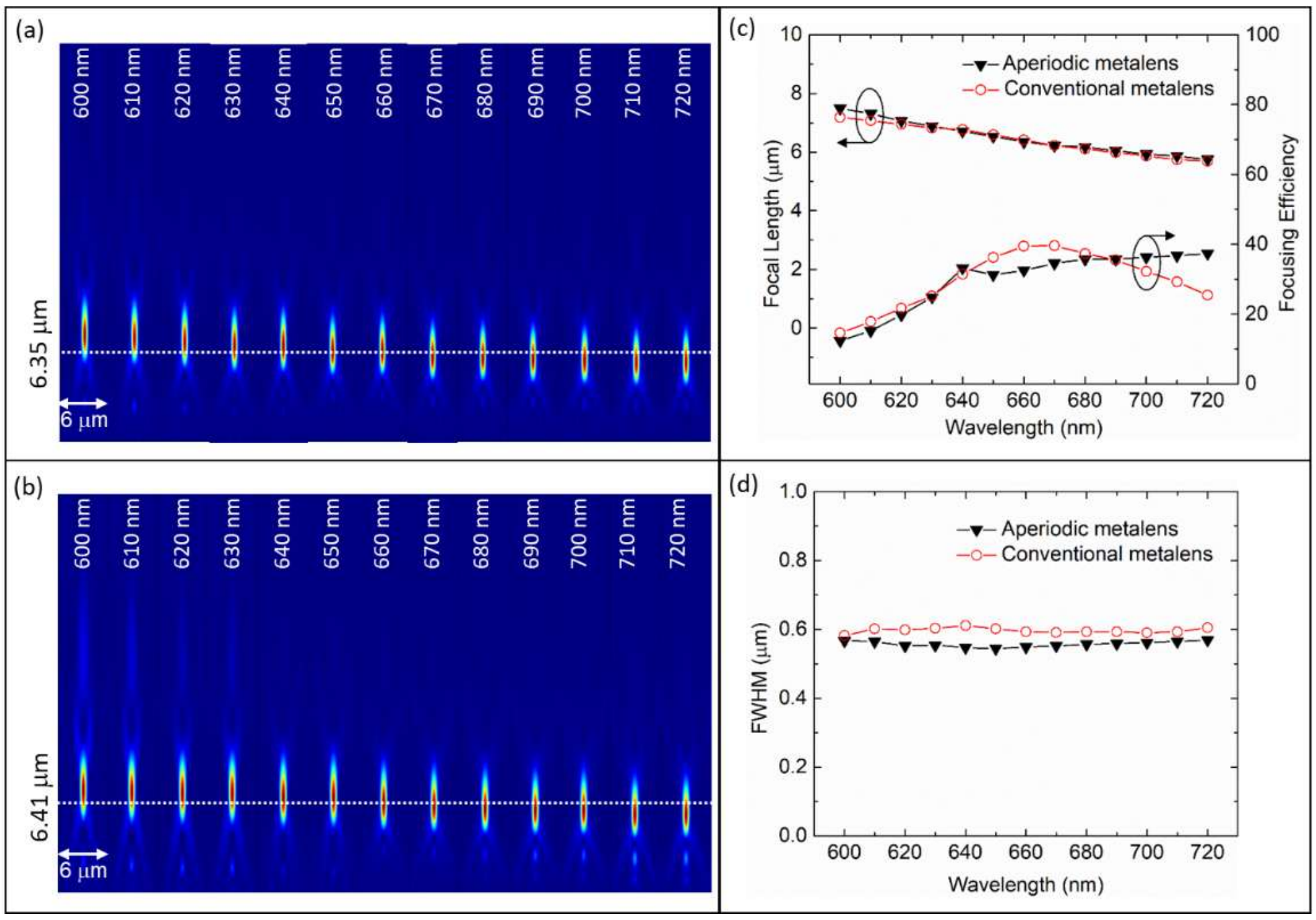

Fig. 8. Chromatic focusing performance analysis for high-NA metalenses. Longitudinal intensity distributions (*xz*-plane) simulated across the 600–720 nm spectral range for (a) the aperiodic metalens and (b) the conventional size-variant design, illustrating the chromatic focal shift. (c) Quantitative comparison of focal length and focusing efficiency as a function of wavelength. (d) Full Width at Half Maximum (FWHM) of the focal spot versus wavelength. The metalenses analyzed here correspond to the monochromatic designs presented in Fig. 7, which were designed for $\lambda_d$=660 nm and are evaluated here under broadband illumination.

For the aperiodic metalens, the focal shift is $\Delta f$=1.74 μm, while for the conventional metalens it is $\Delta f$=1.5 μm; here, both exceed the highly constrained diffraction-limited DoF ($\lambda/NA^2$) of ~1 μm. It is worth noting that in this high-NA regime, the aperiodic design exhibits a slightly larger focal shift than the conventional one. This behavior is consistent with our theoretical discussion: at high NA, the metalens encompasses multiple Fresnel zones, introducing diffractive chromatic aberrations that begin to dominate over the passive dispersive suppression provided by our architecture. Nevertheless, the aperiodic design still maintains a consistently tighter focal spot across the entire spectrum, securing superior spatial resolution in this regime.

While the focal shift behavior appears similar at first glance, Fig. 8(d) reveals the critical distinction: the higher achromatic property observed in the moderate-NA aperiodic design is preserved in the high-NA regime. The aperiodic metalens maintains a consistently sharper focal spot across the entire spectrum (600–720 nm). It outperforms the conventional metalens by exhibiting a smaller average spot width, with an average FWHM of 557 nm compared to 597 nm for the conventional metalens.

The key performance metrics for both high-NA designs are summarized in Table S4 (In Section S8 of Supplemental Document). The data demonstrate similar performance regarding longitudinal chromatic aberration, yet highlight two distinct advantages of the aperiodic paradigm. First, it maintains a consistently tighter focal spot across the operational bandwidth. Second, regarding focusing efficiency, while the conventional design exhibits a higher peak, the average efficiency is identical for both designs (~30%). Crucially, the aperiodic metalens demonstrates higher spectral stability, characterized by a significantly lower efficiency variation (of 8.82% vs. 12.54%).

### *4.3 Metalens with large diameter*

This section evaluates the optical performance of an aperiodic metalens designed with a large diameter. To provide a fair comparison, both the aperiodic and conventional periodic metalenses were designed with identical target optical parameters: $f$=60.0 μm, and an aperture diameter of 35 μm. The proposed aperiodic metalens consists of structurally identical nanorods arranged in 90 concentric rings. Simulations yielded an effective focal length of 59.45μm (NA=0.3) for the aperiodic design, and 58.56 μm (NA=0.3) for the conventional counterpart at the design wavelength (660 nm).

Figure 9 presents their focusing performance over a 120 nm bandwidth (600–720 nm) centered on the design wavelength. The focusing performance shows that the aperiodic metalens produces a tighter focal spot at the design wavelength. Specifically, the aperiodic design achieves a FWHM of 1.209 μm, which is 7.5% larger than its theoretical diffraction limit ($\lambda/2NA$) of 1.125 μm. In contrast, the conventional metalens produces a larger focal spot with a FWHM of 1.230 μm, deviating by 11% from its diffraction limit (1.108 μm). These results provide clear evidence that, even in the large diameter regime, the aperiodic design paradigm effectively enhances focusing resolution, approaching the diffraction limit more closely than the conventional approach.

The chromatic performance of both designs shown in Figs. 9(a) and 9(b) visually suggests similar chromatic focal-shift behavior for both architectures. Quantitatively, the longitudinal chromatic aberration is comparable, as shown in Fig. 9(c); the RMSE of the focal length variation is 4.162 μm for the conventional metalens and 3.805 μm for the aperiodic metalens. Regarding efficiency, the conventional design exhibits higher focusing efficiency at wavelengths below 690 nm, while the aperiodic metalens outperforms it at wavelengths above 690 nm. The focal shift for the aperiodic metalens is $\Delta f$=10.74 μm, and $\Delta f$=11.34 μm for the conventional metalens. As expected for extended apertures, both values fall outside the diffraction-limited DoF ($\lambda/NA^2$) of ~7.3 μm.

While the focal shift behavior appears similar at first glance, Fig. 9(d) reveals again the critical distinction: the higher focusing resolution observed in both the moderate-NA and high-NA aperiodic designs is preserved in the large diameter regime. The aperiodic metalens maintains a consistently sharper focal spot across the entire spectrum (600–720 nm). It outperforms the conventional design by exhibiting a smaller average spot width, with an average FWHM of 1.201 µm compared to 1.234 µm for the traditional metalens. Additionally, the aperiodic metalens exhibits no diffraction orders beneath the focal spot, in contrast to the conventional metalens, where these diffraction orders are clearly observed, as shown in Figs. 9(a) and 9(b).

The key performance metrics for the large-aperture designs are summarized in Table S5 (Section S8 of Supplemental Document). The data demonstrate again a good chromatic performance despite the larger size of the metalens. In addition, to establish the scalability of our theoretical framework, we demonstrate analytical scalability by solving the layout equations (Eqs. 5-7) for a large millimeter-scale aperiodic metalens (2000 rings, $D_m \approx 1$ mm, $f$=1 mm), which is shown in Fig. S6 (Section S9 of Supplemental Document).

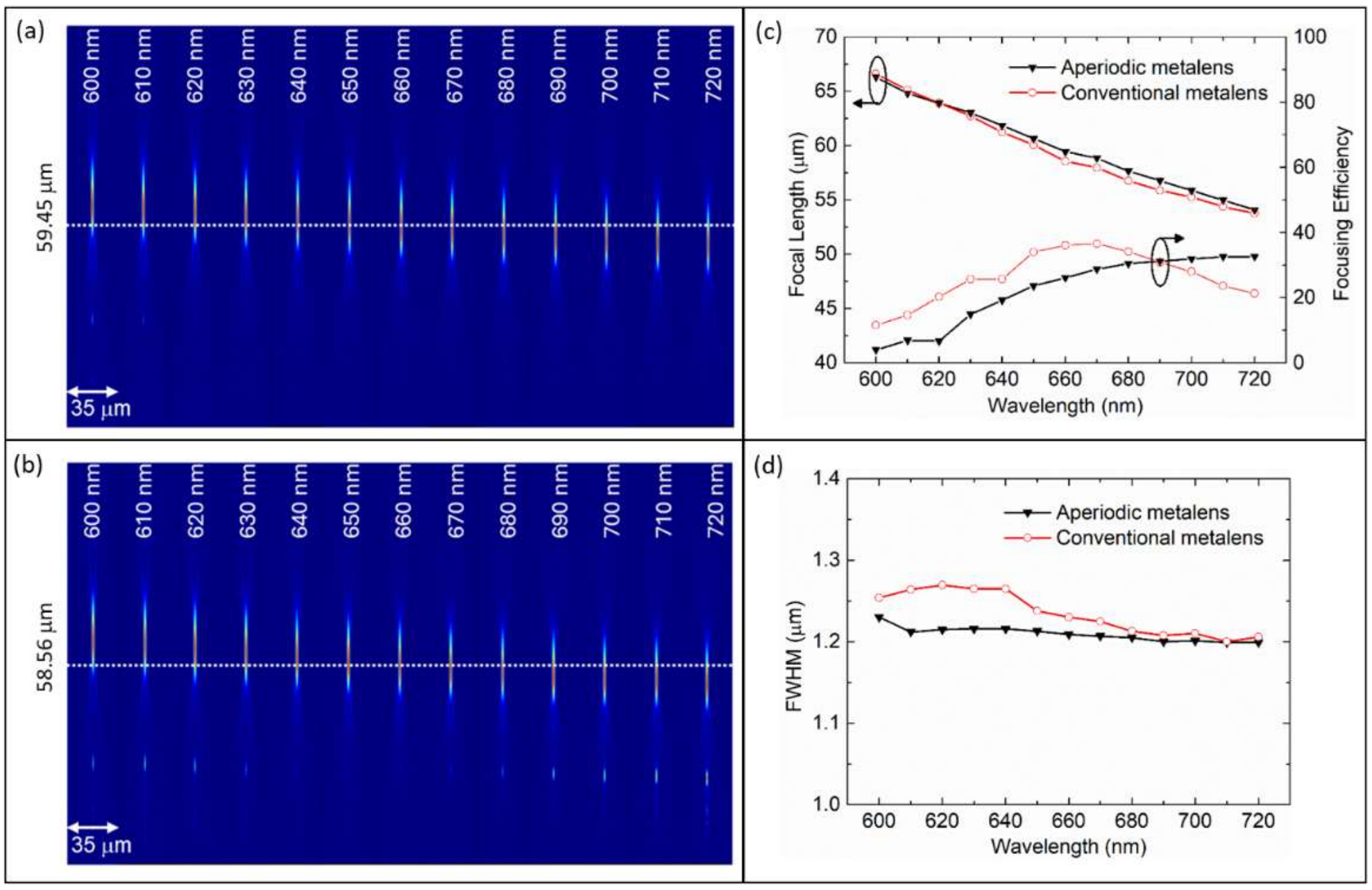

Fig. 9. Chromatic focusing performance analysis for large-diameter metalenses. Both initial designs share identical target optical parameters, with an aperture diameter of 35 µm. Longitudinal intensity distributions (xz-plane) simulated across the 600–720 nm spectral range for (a) the aperiodic metalens and (b) the conventional size-variant design, illustrating the chromatic focal shift. The simulated focal lengths are 58.56 µm for the conventional metalens and 59.45 µm for the aperiodic metalens. (c) Quantitative comparison of focal length and focusing efficiency as a function of wavelength. (d) Full Width at Half Maximum (FWHM) of the focal spot versus wavelength. The aperiodic metalens analyzed here corresponds to the layout shown in Fig. 3(c), designed for $\lambda_d$=660 nm and evaluated under broadband illumination.

### *4.4 Metalens with a different material ($TiO_2$)*

To establish the material generality of our proposed architecture, this section evaluates the optical performance of an aperiodic metalens composed of $TiO_2$ nanorods. For a direct comparison, both the aperiodic and conventional periodic metalenses were designed with the same target optical parameters utilized in Section 4.1: $f$=9.9µm and an aperture diameter of 5.9 µm. The proposed aperiodic metalens consists of structurally identical nanorods arranged in 17 concentric rings. At the design wavelength (660 nm), simulations yielded an effective focal length of 7.13 μm (NA = 0.41) for the aperiodic design, and 7.07 μm (NA = 0.41) for its conventional counterpart.

Figure 10 presents their focusing performance over a 120 nm bandwidth (600–720 nm) centered on the design wavelength. The focusing performance shows that the aperiodic metalens produces a tighter focal spot at the design wavelength. Specifically, the aperiodic design achieves a FWHM of 0.821 μm, which is 2.3% larger than its theoretical diffraction limit ($\lambda/2NA$) of 0.802 μm. In contrast, the conventional metalens produces a larger focal spot with a FWHM of 0.877 μm, deviating by 10.2% from its diffraction limit (0.796 μm). These results provide clear evidence that, even when using a different material, the aperiodic design paradigm effectively enhances focusing resolution, approaching the diffraction limit more closely than the conventional approach.

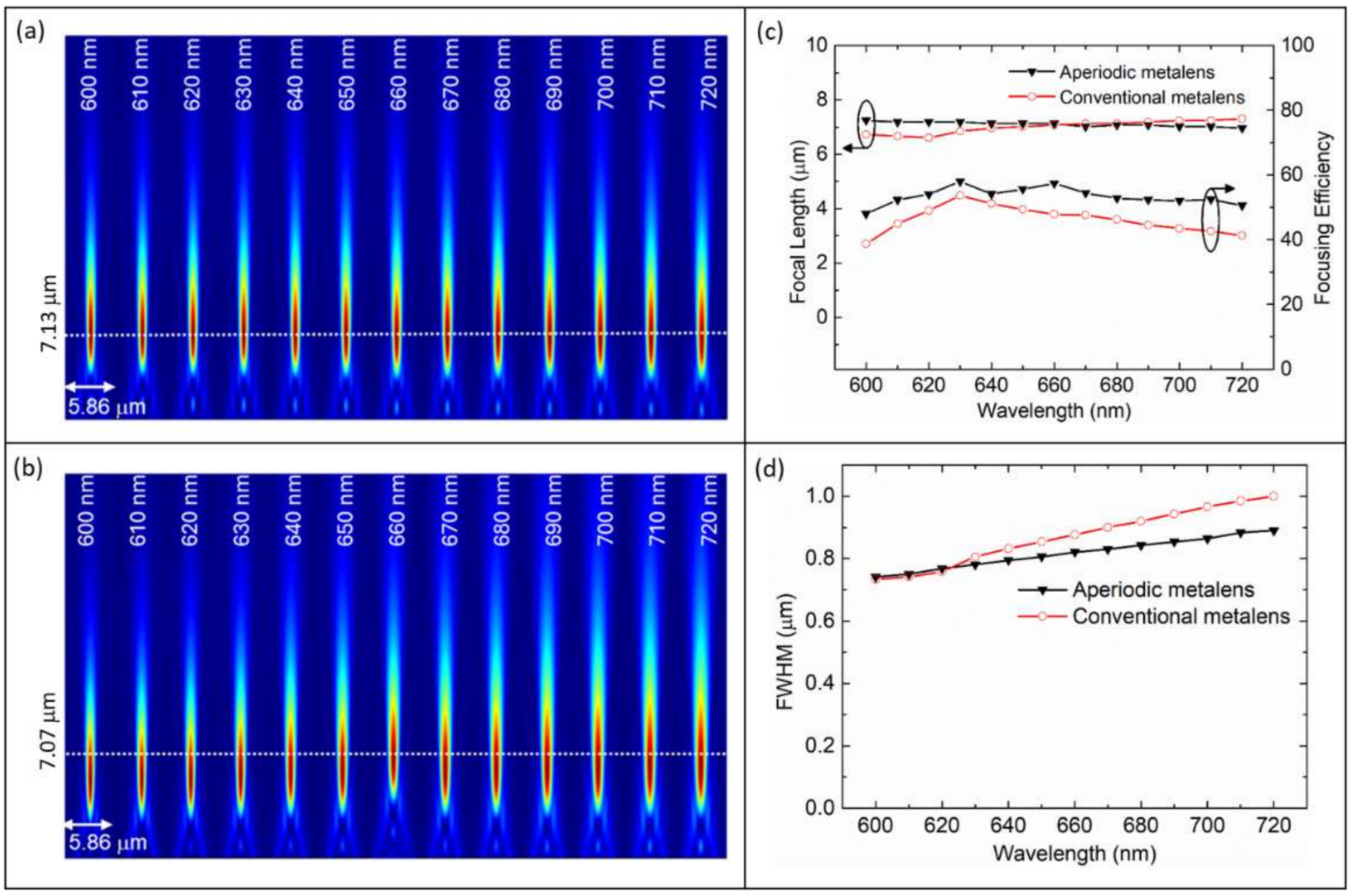

Fig. 10. Chromatic focusing performance analysis for the $TiO_2$ metalens. Both initial designs share identical target optical parameters to those of the first example (Fig. 4). Longitudinal intensity distributions (xz-plane) simulated across the 600–720 nm spectral range for (a) the aperiodic metalens and (b) the conventional size-variant design, illustrating the chromatic focal shift. The simulated focal lengths are 7.07 μm for the conventional metalens and 7.13 μm for the aperiodic metalens. (c) Quantitative comparison of focal length and focusing efficiency as a function of wavelength. (d) Full Width at Half Maximum (FWHM) of the focal spot versus wavelength. The aperiodic metalens analyzed here corresponds to the layout shown in Fig. 3(d), designed for $\lambda_d$=660 nm and evaluated under broadband illumination.

The chromatic performance of both designs shown in Figs. 10(a) and 10(b), visually suggests similar chromatic focal dispersion behaviors for both architectures. Quantitatively, the longitudinal chromatic aberration is substantially reduced, as shown in Fig. 10(c); the RMSE of the focal length is reduced from 0.253 μm for the conventional metalens to just 0.087 μm for the aperiodic metalens, a reduction of nearly 66%. Regarding efficiency, the aperiodic design exhibits higher focusing efficiency at all considered wavelengths. The aperiodic metalens gives a peak focusing efficiency of 58%, a value that could be further increased without changing the metalens design by applying antireflection coatings to the nanopillars [37]. For the $TiO_2$ aperiodic metalens, the focal shift is exceptionally small ($\Delta f$=0.23 μm), outperforming the conventional metalens ($\Delta f$=0.46 μm). Both shifts fall comfortably within the diffraction-limited DoF ($\lambda/NA^2$) of ~3.9 μm.

While the focal shift behavior appears similar at first glance, Fig. 10(d) reveals again the critical distinction: the higher focusing resolution observed in the previous aperiodic designs is preserved in the $TiO_2$ metalens. The aperiodic metalens maintains a consistently sharper focal spot across the entire spectrum (600–720 nm). It outperforms the conventional design by exhibiting a smaller average spot width, with an average FWHM of 0.818 μm compared to 0.871 μm for the conventional metalens.

Crucially, we demonstrate that switching to a low-loss platform ($TiO_2$, which has a negligible extinction coefficient $k$ in the visible spectrum) raises the peak efficiency of the aperiodic metalens to ~58% (up from ~35% in Si), while strictly preserving its intrinsic spectral behavior (Fig. 10). This confirms that the proposed variable-periodicity mechanism survives beyond the high-index Si regime. Furthermore, as shown in Fig. 10(d), the aperiodic metalens maintains a consistently sharper focal spot across the entire spectrum. Overall, this alternative material design systematically matches or surpasses the performance metrics of the conventional design (Table S6, Supplemental Document).

## 5. Discussion

The results presented in Section 4 validate the physical concept underlying the proposed architecture: aperiodic metalenses built from a single invariant nanocylinder inherently provide passive chromatic aberration suppression while maintaining near-diffraction-limited focusing. Unlike conventional designs that rely on complex libraries of varying nanorod diameters to map phase, our fully deterministic approach achieves phase control exclusively by modulating the local lattice periodicity. As established in Section 2.2, the physical mechanism underpinning this performance is the nearly linear dependence of the effective refractive index on the fill factor, which arises from decoupling the diameter-dependent geometric resonances from the waveguide propagation constants. We now discuss some practical implications, scope, and limitations of this design paradigm.

It is important to contextualize the term "achromatic" within the current landscape of metalens research. As discussed in recent reviews [18–20], true broadband achromaticity typically requires independent control over group delay and group delay dispersion to satisfy the time-bandwidth product limit (Eq. 6 in [19]), often necessitating advanced multi-atom unit cells or high-contrast resonators. The aperiodic design presented here does not attempt full dispersion cancellation in that strict sense. Instead, it provides passive suppression of chromatic aberration, a new path toward broadband achromaticity, it provides a simple baseline for dispersion-engineered metasurfaces. By maintaining a constant nanocylinder geometry, the dispersive contribution of the waveguide mode remains spatially uniform, minimizing the higher-order dispersion terms that typically plague periodic designs with varying resonator diameters. Consequently, while the focal shift is not eliminated, it is significantly suppressed compared with the periodic equivalent. Our theoretical and numerical results therefore show that the aperiodic spatial distribution itself can linearize the group-delay response to a useful approximation, providing an intrinsic dispersive compensation mechanism. Importantly, simulations using a low-loss $TiO_2$ platform (Section 4.4) confirm that this mechanism is fundamentally architectural rather than material-specific. Transitioning to a lower-loss material preserved the passive achromatic response while increasing peak efficiency, indicating that the moderate efficiencies obtained with silicon are primarily associated with material absorption rather than with a fundamental limitation of the aperiodic concept. Thus, the observed efficiencies should be viewed as a trade-off expected from a baseline platform that uses a single building block and no geometry optimization, rather than as an upper bound on the architecture.

Our comparative analyses indicate that while the aperiodic design outperforms the conventional one at both moderate (NA=0.4) and high (NA=0.8) numerical apertures, the relative performance gain is less pronounced in the high-NA regime. A first possible explanation would be limited phase-sampling resolution, since phase gradients become steeper at high NA. However, the aperiodic design actually achieves a slightly finer spatial sampling for a full $2\pi$ cycle ($p_{\min}+p_{\max}$ = 570 nm) than the classical design ($2p$=600nm), which rules out sampling resolution as the dominant cause. The more likely explanation is a combination of effects, including partial breakdown of the effective-medium approximation in regions of strong phase gradient, near-field coupling between adjacent waveguides, and material-dependent spectral resonances. In these regimes, local nonlocality can introduce phase errors that the linear effective-index model (Eq. S8) does not fully capture. Furthermore, as our multi-material analysis implies, performance saturation is also influenced by material-dependent spectral resonances. A more fundamental limitation, however, relates to the physics of dispersionless designs. As established by Arbabi et al. [38], designs employing meta-atoms or metamolecules with constant delay (dispersionless) perform ideally only for metalenses comprising a single Fresnel zone (i.e., total phase accumulation $|\Phi|<2\pi$). For high-NA or large-aperture metalenses involving multiple Fresnel zones (concentric diffraction gratings composed of meta-atoms), dispersionless designs inherently incur chromatic aberration due to the inability to compensate for the spectral diffraction across different zones. Nevertheless, as explicitly demonstrated by our large-aperture simulations (Section 4.3), even when scaling up to 90 rings and multiple Fresnel zones, the aperiodic architecture successfully maintains a tighter focal spot (smaller FWHM) than the conventional design. This confirms that the chromatic compensation is robust, highly scalable, and not merely an artifact of increased spherical aberration extending the depth of focus, but rather a genuine improvement in wavefront quality across the spectral range. We therefore interpret the large-aperture and high-NA saturations as a consequence of the increased physical complexity of the focusing problem, rather than as a failure of the aperiodic principle itself. In resume, the observed saturation at high NA and large size is consistent with fundamental physical limits (e.g., Fresnel zone effects) and indicates that the current deterministic framework is a powerful starting point. We anticipate that integrating this aperiodic lattice with inverse-design methods could further push these boundaries.

While the primary aim of this article is to establish the physical principles and theoretical limits of a geometry-invariant aperiodic architecture, a critical practical consideration remains: the fabrication of the densest regions. Satisfying the phase profile in these areas—mainly at the center of the metalens and, to a lesser extent, at the Fresnel zone boundaries—requires lattice periods as small as 150 nm. With a constant nanorod diameter of 120 nm, this leaves interstitial air gaps of roughly 30 nm. Given a structure height of 800 nm for Si (and 1200 nm for $TiO_2$), this corresponds to a high air-gap aspect ratio of approximately 27:1 (and 40:1 for $TiO_2$). Crucially, it must be emphasized that these high aspect ratios apply exclusively to the interstitial voids, not to the structural pillars themselves. The dielectric nanorods maintain a mechanically robust aspect ratio of less than 7:1 (for Si) and 10:1 (for $TiO_2$), ensuring structural stability and preventing capillary collapse during processing. Furthermore, this extreme gap aspect ratio represents a worst-case scenario occurring only between two of the smallest unit cells; across the full phase coverage range, the gap aspect ratio relaxes significantly, dropping to below 3:1 for the largest unit cell (420 nm). Accordingly, while the fabrication of the densest regions represents a challenge for standard nanolithography [39], the proposed architecture establishes an analytically tractable and deterministic physical baseline. Future research can leverage advanced process optimization to achieve experimental realization, as similar high-aspect-ratio gaps have already been successfully demonstrated in recent literature [39]. Additionally, at the minimum gap between nanorods (30 nm), near-field crosstalk between adjacent elements could theoretically disrupt the local phase approximation. To rigorously verify the validity of our phase library in this dense aperiodic environment, we compared the phase response of an isolated unit cell against arrayed unit cells via FDTD simulations. The simulated phase and transmission responses remain broadly consistent even as neighboring nanocylinders are introduced, confirming that inter-element coupling does not substantially alter the local optical response under our design conditions (Section S10, Supplemental Document).

Beyond the fundamental physics of chromatic suppression, the radical reduction of geometric diversity—achieved by using a single, invariant building block—could have intriguing implications for fabrication tolerance. Previous studies on dielectric metalenses have demonstrated that typical fabrication deviations, such as variations in pillar height, diameter, or sidewall profile, measurably degrade focusing efficiency and focal spot quality [40–42]. The proposed aperiodic architecture, however, offers a potential intrinsic resilience to systematic fabrication errors. For example, a uniform over-etch or under-etch across the metasurface would shift the phase response of all structurally identical pillars by an approximately equal amount. Because focusing is governed by the spatial phase gradient ($\partial\Phi/\partial r$), adding a global, uniform phase offset inherently preserves the focal behavior. In contrast, applying a uniform geometric deviation to a conventional library-based metalens affects its varied pillar diameters disproportionately. This non-linear response introduces spatially non-uniform phase errors across the aperture, disrupting the intended phase gradient. While this qualitative mechanism points toward a highly favorable tolerance profile, rigorous quantitative validation—such as Monte Carlo analyses of dimensional deviations and placement disorder—remains an important avenue for future work. Exploring this robustness is particularly compelling in the context of the broader literature on engineered disorder in photonic structures [15,43–45].

Regarding the coordinate-generation algorithm presented in Section 3, we deliberately adopted a strict radial discretization and a locally square unit-cell topology to establish a mathematically transparent, analytically tractable baseline—demonstrating that the aperiodic principle succeeds even without complex topological optimization. As seen in Fig. 3, strictly enforcing this unit-cell 'squareness' condition inherently introduces a localized azimuthal discontinuity, visible as a boundary discontinuity between the first ($j$=0) and last ($j=M_i$) angular coordinates ($\theta_{ij}$) of each concentric ring. Although this azimuthal discontinuity is localized at the angular boundary of each ring, its quantitative effect on focusing efficiency and aberrations has not been independently isolated; however, the present simulations demonstrate that the overall architecture retains its focusing functionality. Crucially, the relative influence of this boundary effect naturally diminishes as the metalens aperture increases, since the number of affected unit cells scales much more slowly (linearly with the number of rings) than the total number of meta-atoms across the aperture (which scales quadratically with the diameter). Nevertheless, this foundational algorithm opens inspiring avenues for future geometric and algorithmic optimization. Because the aperiodic layout naturally resembles a spiral architecture, future coordinate-generation schemes could eliminate the azimuthal discontinuity entirely by employing continuous spiral trajectories, or simply by staggering the angular starting locations of each ring with minor local adjustments. Furthermore, transitioning from locally square to hexagonal unit-cell tessellations represents another highly promising direction; hexagonal arrays naturally provide a superior packing density (fill ratio) and higher spatial symmetry, which could facilitate experimental nanoscale fabrication while pushing the focusing efficiency and chromatic performance of aperiodic metalenses even further.

We conclude this section by emphasizing that the principal contribution of this work is the establishment of a deterministic physical framework that decouples phase from geometry. This framework provides a transparent and analytically tractable baseline for future metasurface designs that may combine aperiodicity with inverse optimization or other advanced design strategies. By providing a fully reproducible coordinate-generation methodology and isolating the local period as the sole degree of freedom, we offer the metasurface community a new foundational baseline, one that shifts the design paradigm from geometry-variant libraries to geometry-invariant, periodicity-engineered architectures.

## 6. Conclusion

This work establishes and numerically validates a paradigm shift in metalens design, fundamentally decoupling phase modulation from meta-atom geometry. By shifting the design philosophy from geometry-variant libraries to geometry-invariant, positionally-engineered architectures, we have demonstrated that full $2\pi$ phase coverage can be achieved exclusively through local periodicity modulation. Furthermore, this architecture intrinsically approximates the ideal achromaticity condition more closely than conventional periodic designs, driven by the linear scaling of the effective refractive index with the local fill factor. Unlike conventional approaches that rely on complex multi-parameter libraries, our method provides a fully deterministic, analytically tractable framework. Supported by our open-source coordinate-generation scripts (Python/MATLAB codes) and based on a single nanostructural building block, this approach is inherently reproducible and scalable. Full-wave numerical simulations rigorously validate our theoretical predictions, confirming the viability of passive chromatic aberration suppression across the visible spectrum.

Our broadband simulations confirm that this mechanism yields an intrinsic level of passive achromatic performance. For a moderate numerical aperture (NA=0.4), the aperiodic design reduced longitudinal chromatic aberration by nearly 42% (Si) and 66% ($TiO_2$) while simultaneously enhancing focusing resolution to produce a tighter, near diffraction-limited spot across the entire visible spectral range. Even in the challenging high-NA regime (NA=0.8), where traditional effective medium theories are strained and multiple Fresnel zones introduce fundamental chromatic aberrations, the aperiodic approach maintained high wavefront quality, consistently outperforming its size-variant counterpart. Furthermore, we validated the robust scalability of this framework through large-aperture full-wave simulations (~35 μm) and analytical macro-scale layouts (up to 1 mm), confirming that the enhanced focal confinement is preserved at scale. Crucially, by evaluating the architecture with a low-loss $TiO_2$ platform, the peak focusing efficiency increased while strictly preserving the intrinsic passive achromaticity, confirming that the mechanism is architectural rather than material-specific.

Consequently, the proposed framework serves as a foundational, deterministic platform for a new class of scalable, polarization-insensitive flat optics. By providing a reproducible, analytically tractable, and single-building-block approach, we offer a robust baseline upon which future experimental and computational advances can be built. Future efforts may focus on integrating this aperiodic lattice strategy with inverse-design methods to further optimize both broadband performance and efficiency. Future work should address experimental validation, fabrication tolerances, and practical implementation challenges, particularly the formation of narrow interstitial gaps with high aspect ratios. Alternative fabrication strategies may also be explored. Additionally, this paradigm can be extended to other spectral ranges and diverse meta-atom types [46]. Ultimately, this work opens a new direction in meta-optics, moving beyond the traditional reliance on geometry-variant libraries toward aperiodic, geometry-invariant designs.

## Disclosures

The authors declare that there are no conflicts of interest related to this article.

## Data availability

Data underlying the results presented in this paper are not publicly available at this time but may be obtained from the authors upon reasonable request.

## Supplemental document

See Supplement 1 for supporting content.

Supplemental Document for

# Aperiodic metalenses with identical nanocylinders: positional phase control and passive chromatic aberration suppression

Ivan Moreno, J. Carlos Basilio-Ortiz

Unidad Académica de Ciencia y Tecnología de la Luz y la Materia, Universidad Autonoma de Zacatecas, 98060 Zacatecas, Mexico

IM e-mail: imorenoh@uaz.edu.mx

JCBO e-mail: jc_atlet_3000@hotmail.com

## S1. A condition for achromatic focusing in metalenses

Chromatic aberration in metalenses manifests as a frequency-dependent shift of the focal length, originating from the combined effects of material dispersion, resonant scattering, and the intrinsically diffractive nature of metasurfaces. In propagation-phase metalenses, dispersion plays a dominant role and is commonly analyzed by expanding the required phase profile in a Taylor series around the design frequency, where the zeroth-order term corresponds to the phase at the design frequency and the first-order term represents the group delay.

Within the propagation phase mechanism, the phase profile can be written as

$$\Phi(r,\omega) \approx \frac{\omega}{c} n_{eff}(r,\omega)\, H \ , \tag{S1}$$

where $\omega$ is the angular frequency, $c$ is the speed of light, $H$ is the nanorod height, $n_{\mathrm{eff}}(r,\omega)$ is the effective refractive index of the unit cell, and $r$ denotes the radial distance from the metalens center [5]. The corresponding group delay, defined as the derivative of the phase with respect to frequency and evaluated at the design frequency $\omega_d$, is given by

$$\frac{\partial}{\partial\omega}\Phi(r,\omega)\Big|_{\omega=\omega_d} \approx \frac{1}{c}\left[n_{eff}(r,\omega_d) + \omega_d \frac{\partial}{\partial\omega} n_{eff}(r,\omega)\Big|_{\omega=\omega_d}\right] H \tag{S2}$$

Conventional achromatic metalens designs focus on engineering this group delay and its higher-order dispersion. However, the position of the focal point is not determined by the absolute phase profile, but by the radial phase gradient, which defines the local transverse wavevector $k_r$ that is responsible for focusing.

We therefore derive a condition for achromatic focusing based on the spatial phase gradient. This derivation relies on the generalized law of refraction for optical metasurfaces with a radially varying phase profile $\Phi(r)$, given by [4]

$$\frac{\mathrm{c}}{\omega}\frac{\mathrm{d}\Phi}{\mathrm{d}r} = n_{\mathrm{t}} sin\theta_{\mathrm{t}} - n_{\mathrm{i}} sin\theta_{\mathrm{i}} \ , \tag{S3}$$

where $\theta_{\mathrm{i}}$, $\theta_{\mathrm{t}}$ are the angles of incidence and transmission, respectively, and $n_{\mathrm{i}}$ and $n_{\mathrm{t}}$ are the refractive indices of the incident and transmitted media.

For a standard focusing metalens (Fig. S1(a)), illuminated under normal incidence ($\theta_{\mathrm{i}}$=0) and focusing into air ($n_{\mathrm{t}}$=1), Eq. (S3) reduces to

$$\frac{\mathrm{c}}{\omega}\frac{\mathrm{d}\Phi(r)}{\mathrm{d}r} = sin\theta_{\mathrm{t}} \ , \tag{S4}$$

This equation indicates that the phase gradient dependence with $r$ determines the angles $\theta_{\mathrm{t}}$ to maintain $f$ constant. Maintaining a constant $f$ for every $r$ is equivalent to maintaining it for every angle $\theta_{\mathrm{t}}$ (since each $\theta_{\mathrm{t}}$ corresponds to a specific $r$). In an achromatic focusing metalens, the focal length $f$ must remain constant across the entire spectral range of interest $\Delta\omega$ and for all radial positions $r$ or angles $\theta_{\mathrm{t}}$. Consequently, to achieve achromatic focusing, Eq. (S4) becomes an achromatic focusing condition by making it frequency-independent. Specifically, by making $\sin\theta_{\mathrm{t}}=(c/\omega)\mathrm{d}\Phi/\mathrm{d}r=(c/\omega_d)\mathrm{d}\Phi/\mathrm{d}r|_{\omega=\omega d}$. This imposes the following achromaticity condition:

$$\frac{\partial}{\partial r}\Phi(r,\omega) = \frac{\omega}{\omega_d}\frac{\partial}{\partial r}\Phi(r,\omega_d) \tag{S5}$$

where $\omega$ is an arbitrary frequency, and $\omega_d$ is the design frequency.

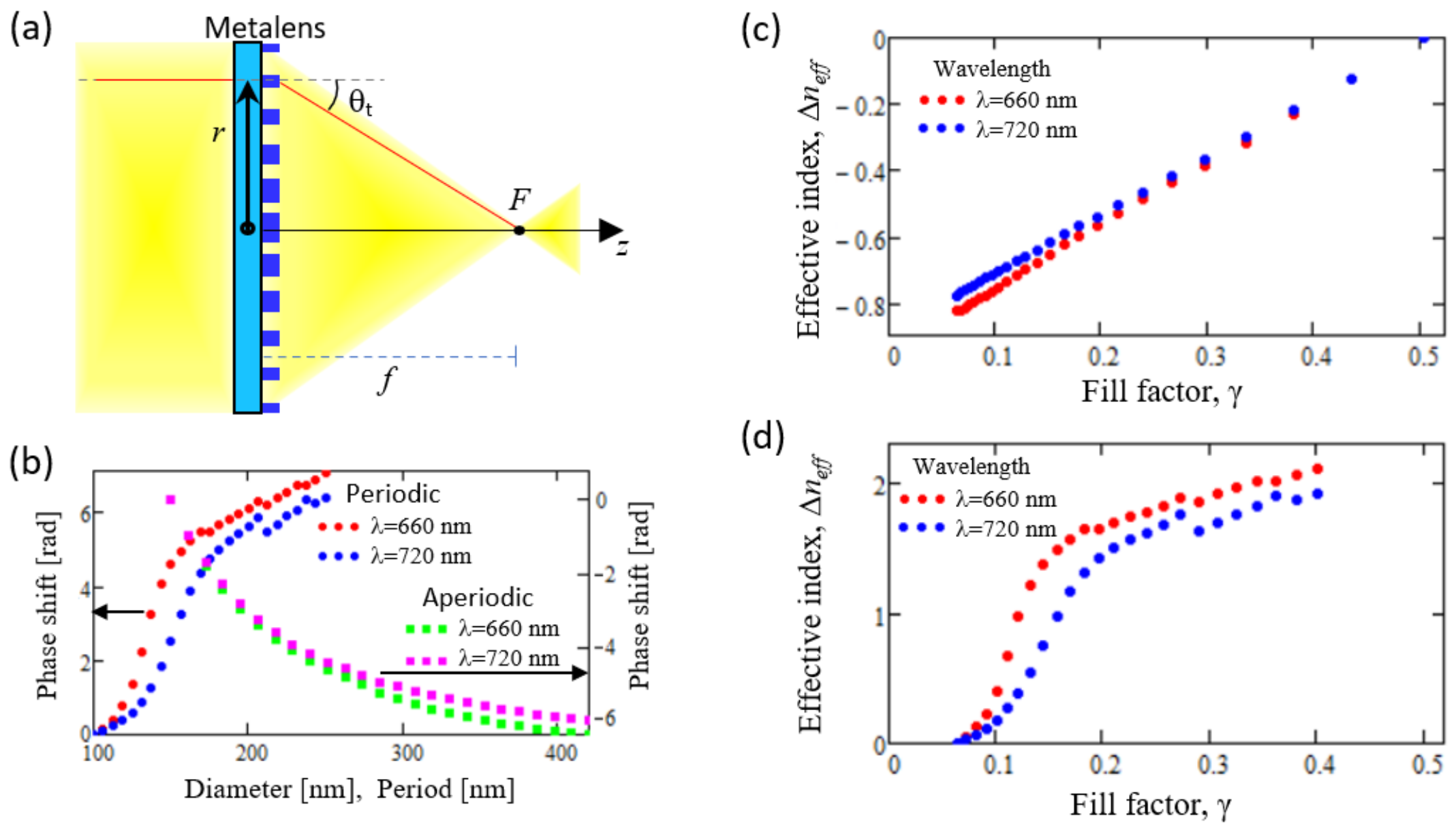

Fig S1. (a) Schematic illustration of a focusing metalens concentrating incident light at a focal distance *f*. Parallel light rays (red lines) are converged by the metalens towards the focal point *F*. (b) FDTD-simulated meta-atom phase shift (Φ) as a function of the nanorod diameter (for the periodic design), and as a function of the unit-cell period (for the aperiodic design). (c,d) FDTD-calculated effective refractive index change ($\Delta n_{eff}$) as a function of the unit-cell fill factor ($\gamma$), where (c) is for the proposed aperiodic meta-atoms (variable period), and (d) for the classical periodic meta-atoms (fixed period). Note the intrinsic linear dependence in the aperiodic approach (c), in contrast to the classical non-linear behavior of the periodic design in (d). Panels (b)-(d) show results for the analyzed Si meta-atoms.

We now evaluate both the proposed aperiodic metalens (variable period, fixed nanopillar) and the classical periodic metalens (fixed period, variable nanopillar) against the achromaticity condition of Eq. (S5). As illustrated in Fig. S1(b), the aperiodic meta-atoms exhibit a smooth phase variation with respect to the unit-cell period, in contrast to the periodic design, which shows a stronger variation as a function of the nanorod diameter. In both architectures, the phase shift is governed by the truncated waveguide approximation (Eq. (S1)), where the accumulated phase depends on the effective index $n_{eff}$.

In the propagation phase mechanism, $n_{eff}$ arises from the hybridization of the optical properties of the nanorod and those of the surrounding medium. The unit cell behaves as an effective medium whose refractive index is modulated by the fill factor γ, which controls light confinement. The extent to which the pillar fills the unit cell is referred to as the fill factor, which is defined as the ratio between the nanopillar volume and the unit-cell volume, $\gamma = V_{pillar}/V_{cell}$. According to effective medium theory [22-25], $n_{eff}$ increases monotonically with $\gamma$. For $\gamma \rightarrow 0$, $n_{eff}$ approaches the index of the background (air), while for $\gamma \rightarrow 1$, it approaches the refractive index of the nanorod material.

For a metalens, the fill factor $\gamma$ is modulated as a function of the radial distance $r$. Therefore, the effective index can be expressed as $n_{eff}(r,\omega) = n_{eff}[\gamma(r),\omega]$. Substituting this into the phase gradient requires applying the chain rule:

$$\frac{\partial}{\partial r}\Phi(r,\omega) = \approx \frac{\omega H}{c}\frac{\partial}{\partial r}\, n_{eff}(r,\omega) = \frac{\omega H}{c}\frac{\partial}{\partial \gamma}\, n_{eff}(\gamma,\omega)\, \frac{\mathrm{d}}{\mathrm{d}r}\, \gamma(r) \tag{S6}$$

By using Eq. (S6) we can check how much a propagation phase metalens meets the ideal achromaticity condition (Eq. S5). By comparing the phase gradient at an arbitrary $\omega$ to that at $\omega_d$. This comparison implies that the metalens achieves ideal achromatic focusing if the derivative of the effective refractive index with respect to the fill factor is invariant to frequency:

$$\frac{\partial}{\partial \gamma}\, n_{eff}(\gamma,\omega) = \frac{\partial}{\partial \gamma}\, n_{eff}(\gamma,\omega_d) \tag{S7}$$

Evaluating this achromaticity condition (Eq. (S7)) requires precise knowledge of $n_{eff}(\gamma,\omega)$, which we extracted via FDTD simulations for both unit cell types (Figs. S1b and S1c). Figs S1(b) and S1(c) show $\Delta n_{\rm eff}$ because the effective index can be written as $n_{\rm eff}[\gamma(r),\omega]=\Delta n_{\rm eff}[\gamma(r),\omega]+n_{\rm o}(\omega)$, where the index shift $\Delta n_{eff}$ provides the required phase shift coverage of the metalens. A key finding is that the aperiodic unit cell exhibits a highly linear dependence of $n_{eff}$ on the fill factor $\gamma$. This linearity allows us to model the effective index as:

$$n_{eff}(\gamma,\omega)\approx\sqrt{n(\omega)}[\gamma(r)-\gamma_0]+n_0(\omega) \tag{S8}$$

where $n(\omega)$ is the refractive index of the pillar, $\gamma_0$ is a reference fill factor (here $\gamma_0$=0.503), and $n_0$ is a reference refractive index. Using this linear approximation, the derivative $\partial n_{eff}/\partial\gamma$ becomes independent of $\gamma$ (and thus independent of $r$), i.e., $\partial n_{eff}/\partial\gamma \approx\sqrt{n(\omega)}$. Consequently, the ratio of derivatives depends only on the material dispersion, not on the local geometry. Therefore, the phase gradient for an arbitrary frequency $\omega$ in terms of the phase gradient at $\omega_d$ for the aperiodic metalens is:

$$\frac{\partial}{\partial r}\Phi(r,\omega)\approx\frac{\omega}{\omega_d}\left(\sqrt{\frac{n(\omega)}{n(\omega_d)}}\right)\frac{\partial}{\partial r}\Phi(r,\omega_d) \tag{S9}$$

This equation deviates from the ideal achromatic condition (Eq. (S5)) only by the ratio $[n(\omega)/n(\omega_d)]^{0.5}$, which is independent of $r$. Our analysis shows that the aperiodic metalens satisfies the ideal achromatic focusing condition to within ~99% across the considered spectral range in the visible ($\Delta\lambda$=120 nm). Specifically, the deviation is only 1.4% at $\lambda$=600 nm, and 1.0% at $\lambda$=720 nm relative to the design wavelength $\lambda_d$=660 nm.

On the other hand, the classical periodic metalens exhibits a nonlinear $n_{eff}(\gamma,\omega)$ response, causing its derivative $\partial n_{eff}/\partial\gamma$ to vary significantly with both $\omega$ and $r$, i.e., $\partial n_{eff}/\partial\gamma\approx F(\gamma(r),\omega)$. The ratio of these derivatives determines the metalens' deviation from the ideal achromatic condition (Eq. (S5)), in particular, how much it deviates from unity. We numerically quantified this deviation $\Delta$ using a quadratic average difference:

$$\Delta=\frac{1}{N}\sum_i^N\sqrt{\left[1-\left.\frac{\frac{\partial}{\partial\gamma}n_{eff}(\gamma,\omega)}{\frac{\partial}{\partial\gamma}n_{eff}(\gamma,\omega_d)}\right|_{\gamma_i}\right]^2} \tag{S10}$$

where $N$ is the number of points considered in the computation. The derivatives in Eq. (S10) are computed from the effective indices for the periodic unit cell (Fig. 1S(c)). Our analysis shows that the periodic metalens satisfies the ideal achromatic focusing condition to within ~30% across the considered spectral range in the visible ($\Delta\lambda$=120 nm). Specifically, the deviation is 80% at $\lambda$=600 nm, and 60% at $\lambda$=720 nm relative to the design wavelength $\lambda_d$=660 nm.

Therefore, the periodic design intrinsically fails to meet the ideal achromaticity condition, showing average deviations of $\Delta\approx$70% over the considered visible spectral range (for Si meta-atoms). In contrast, the intrinsic linearity of the effective index in the aperiodic design (due to the geometric invariance of the nanorod diameter) grants it passive chromatic aberration suppression, showing average deviations of only $\Delta\approx$1% across the same spectral range (for Si meta-atoms).

## S2. Calculation of the period profile for a hyperbolic metalens

To calculate the required unit cell period $p$ as a function of distance $r$, first a relationship between period and phase supplemented by Figs. 2(b) and 2(c) is required. In other words, the data of these figures is employed to fit a function of periods or pitches vs phase shift introduced. We fitted data to the following exponential function:

$$p(r)=ae^{b\Phi}+ce^{d\Phi} \tag{S11}$$

where the coefficients $a$-$d$, for Si meta-atoms, are given by: $a$=0.0983, $b$=–1.1576, $c$=147.4215, and $d$= –0.0961. And the coefficients $a$-$d$, for $TiO_2$ meta-atoms, are given by: $a$=0.1302, $b$= –1.1225, $c$=146.5, and $d$= –0.09335. The Eq. (S11) fits well the data obtained from the FDTD simulations presented in Fig. 2 for the aperiodic meta-atoms. These are displayed in Fig. S2 for comparison purposes.

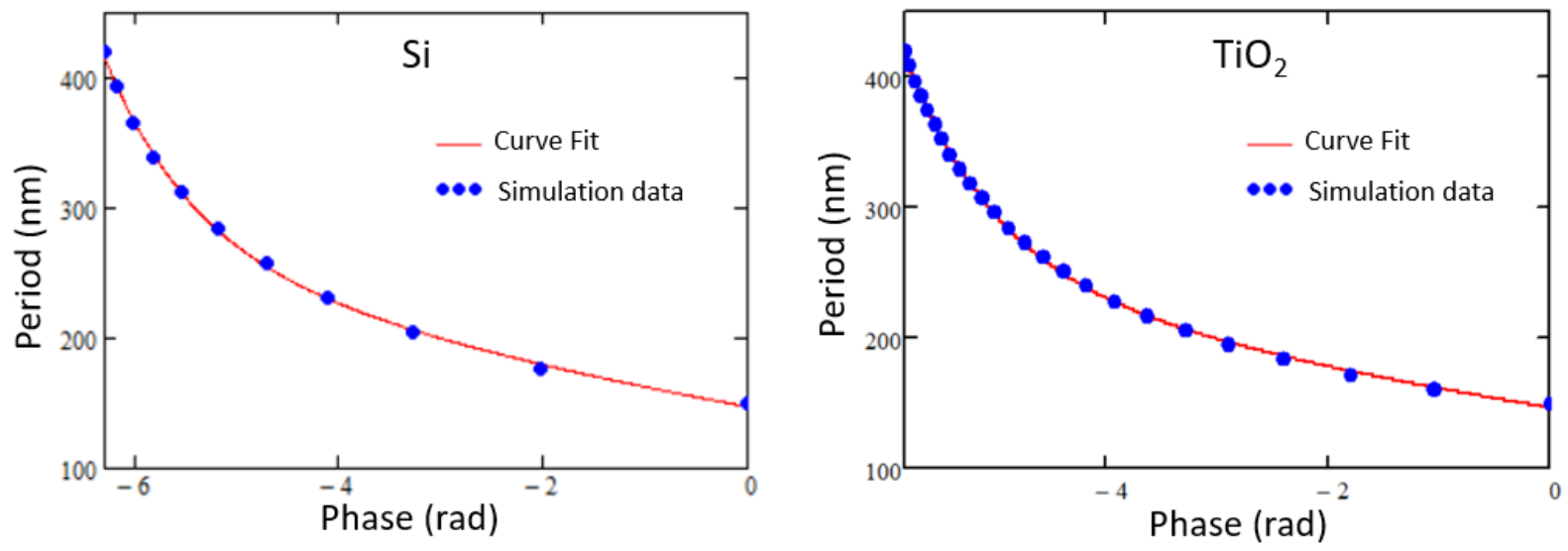

Fig S2. Curve fits (Eq. S12) and simulated data of the unit-cell period versus phase shift for the aperiodic meta-atoms, corresponding to the data shown in Figs. 2(b) and 2(c).

In order to make the aperiodic metalens with the hyperbolic phase shift profile $\Phi(r)$ given by Eq. (4), a spatial distribution of periods dictated by the hyperbolic profile is required. For this purpose, we introduce the hyperbolic profile $\Phi$(r) into Eq. (S11). This gives the desired period distribution $p$(r). For example, for the considered Si meta-atoms it is given by:

$$p(r) = 0.0983e^{-1.1576\Phi(r)} + 147.4215e^{-0.0961\Phi(r)} \qquad \text{(S12)}$$

where $\Phi(r)$ is given by:

$$\Phi(r) = -\frac{2\pi}{\lambda_d}\left(\sqrt{r^2+f^2} - f\right) \qquad \text{(S13)}$$

where $\lambda_d$ is the design wavelength, $f$ is the design focal length, and $r$ is the radial distance from any position on the metalens to its center. For example, we used $\lambda_d$=660nm and $f$=9.9 µm for the analyzed metalens in Section 3.2.1. One can note that the simulated metalens had a smaller focal length for all designs, however it is common that the design focal length and the simulated one (by FDTD method) are not exactly the same.

## S3. Calculations of periods and coordinates of nanorods in a metalens with unit cell distribution in a simple radial mosaic

The metalens surface is divided into concentric rings, and each ring is subdivided into segments acting as individual unit cells (Fig. 2g). The area of these segments is given by

$$A_i = \frac{\Delta\theta_i}{2}\left({\rho_i}^2 - {\rho_{i-1}}^2\right) \quad , \qquad \text{(S14)}$$

where $\rho_i$ and $\rho_{i-1}$ are the radial distances enclosing the $i$-th ring segment (see Fig. 2h), and $\Delta\theta_i$ is the angle subtended by each ring segment. Each ring is partitioned in equal segments so that $\Delta\theta_i$ is constant for each ring. The area of the segment can be written as:

$$A_i = \frac{\Delta\theta_i}{2}(\rho_i + \rho_{i-1})(\rho_i - \rho_{i-1}) \quad , \qquad \text{(S15)}$$

Each ring is partitioned in equal angular intervals, each one of size $\Delta\theta_i$. Depending on the size of $\Delta\theta_i$ , these segments approximate a rectangular or square shape as the radial distance $r$ increases. In order to make these segments become squares as $r$ increase, we can impose a condition to make these segments to be squares as much as possible. Such a "squareness" condition makes the two sides of the ring segment to have the same length. The radial side of the segment has a length $\Delta\rho_i=\rho_i-\rho_{i-1}$, while the angular side has an average length $\Delta s_i = \Delta\theta_i\, r_i = \Delta\theta_i\, (\rho_i+\rho_{i-1})/2$ as shown in Fig. 2(h). Therefore, the "squareness" condition makes $\Delta\rho_i = \Delta s_i$, which is

$$\Delta\rho_i = \frac{\Delta\theta_i}{2}(\rho_i + \rho_{i-1}) = (\rho_i - \rho_{i-1}) \tag{S16}$$

Then we can rewrite the area given by Eq. (S15) as

$$A_i = (\rho_i - \rho_{i-1})^2 \quad , \tag{S17}$$

Now we can equal this area to the area of a squared unit cell with area $p^2$, where $p$ is the unit cell period. In other words, $A_i=p_i^2$, which leads to solve the following nonlinear equation for each ring:

$$\rho_i - \rho_{i-1} - p(r_i) = 0 \quad , \tag{S18}$$

where $p(r_i)$ is the period for a meta-atom at a radial distance $r_i$ (see Fig. 2h), where $i$=1…$N$. Eq. (S18) is solved by substituting in the equation $r_i=(\rho_i+\rho_{i-1})/2$, and then solving recursively for $\rho_i$, for $i$=1…$N$. The central unit cell ($r$=0) was considered to be a circle with area equal to the area of the smallest period for squared unit cells $p_{\min}$, i.e. $\pi\rho_0^2 = p_{\min}^2$. The final solutions are the required radial coordinates $r_i$ for each nanorod of the metalens, which are given by

$$r_i = (\rho_i + \rho_{i-1})/2 \quad , \tag{S19}$$

And the angular coordinate of each nanorod can be obtained from the angular intervals $\Delta\theta_i$. These angular intervals can be obtained by equating the areas of ring segments and squares, i.e. by

$$A_i = \frac{\Delta\theta_i}{2}(\rho_i + \rho_{i-1})(\rho_i - \rho_{i-1}) = [p(r_i)]^2 \tag{S20}$$

where if using Eqs. (S18) and (S19), the angular size is $\Delta\theta_i = p(r_i)/r_i$. Therefore, the polar coordinates of each nanorod are given by $(r_i, \theta_{ij})$, where the angular coordinate is $\theta_{ij}=j\Delta\theta_i$ or

$$\theta_{ij} = j\frac{p(r_i)}{r_i} \quad , \tag{S21}$$

Here $j$ index is associated to the $j$-th segment of each ring, where $j$=0…$M_i$, and then $\theta_{i0}$=0. However, since the full circumference is not always an integer multiple of the angular pitch $\Delta\theta_i$, the number of segments in each ring is determined by rounding to the nearest integer: $M_i$ =Round(360º/$\Delta\theta_i$). Consequently, this fixed discretization introduces a slight azimuthal discontinuity between the last ($\theta_{i,M}$) and first ($\theta_{i,0}$) nanorods in each ring. This artifact arises from strictly meeting the unit-cell squareness condition, which fixes the angular interval size. Such an inhomogeneity (cut) likely introduces a slight decrease in the focusing efficiency of the metalenses. However, this parasitic effect reduces proportionally as the metalens diameter increases; the localized disruption scales linearly with the diameter, whereas the total focused energy scales quadratically with the diameter. While these minor localized lattice disruptions break perfect angular periodicity, mitigating them represents a compelling avenue for algorithmic optimization in future research.

In order to solve Eq. (S18) for the hyperbolic metalens, the targeted $p(r_i)$ must be used. In our case, we simulated the metalens with the materials and parameters that lead to the phase profile, for example Eq. (S12) for Si nanorods. Therefore, by substituting Eq. (S12) into Eq. (S18) gives the particular equations to solve iteratively:

$$\rho_i - \rho_{i-1} - 0.0983e^{-1.1576\Phi[(\rho_i+\rho_{i-1})/2]} + 147.4215e^{-0.0961\Phi[(\rho_i+\rho_{i-1})/2]} = 0 \tag{S22}$$

The initial ring radius $\rho_0$ was given by the smallest period for squared unit cells $p_{\min}$=150nm, shown in Fig. 2(b), which by equating unit cell areas it becomes $\rho_0$=84.63nm.

## S4. Spatial coordinates of nanorods

Section 4 analyzed four aperiodic metalens designs, the first two cases were: one with a moderate numerical aperture (NA=0.42) and the other with a high numerical aperture (NA=0.8). This section provides the radial coordinates of the nanorods ($r_i$) along with the parameters necessary to compute their angular coordinates ($\theta_{ij}$). Table S1 summarizes the structural data for the NA=0.4 metalens, while Table S2 presents the corresponding data for the NA=0.8 design. In both cases, the angular coordinates are calculated using the relation $\theta_{ij}=j\Delta\theta_i$, where $j$ is an integer index ranging from 0 to $M_i$ ( $j$=0…$M_i$).

Table S1. Spatial coordinates of nanorods for the aperiodic metalens with NA=0.42.

| $i$ [$i$-th ring] | $r_i$ [nm] | $\Delta\theta_i$ [rad] | $M_i$ |
|---|---|---|---|
| 0 | 0 | 0 | 0 |
| 1 | 158.5 | 0.932 | 6 |
| 2 | 306.4 | 0.484 | 12 |
| 3 | 455.0 | 0.327 | 18 |
| 4 | 604.5 | 0.248 | 24 |
| 5 | 755.2 | 0.201 | 30 |
| 6 | 907.6 | 0.169 | 36 |
| 7 | 1062 | 0.146 | 42 |
| 8 | 1219 | 0.13 | 47 |
| 9 | 1378 | 0.117 | 53 |
| 10 | 1541 | 0.107 | 58 |
| 11 | 1708 | 0.099 | 62 |
| 12 | 1880 | 0.093 | 67 |
| 13 | 2057 | 0.087 | 71 |
| 14 | 2240 | 0.083 | 74 |
| 15 | 2431 | 0.08 | 77 |
| 16 | 2632 | 0.078 | 79 |
| 17 | 2846 | 0.078 | 80 |

Table S2. Spatial coordinates of nanorods the aperiodic metalens with NA=0.8.

| $i$ [$i$-th ring] | $r_i$ [nm] | $\Delta\theta_i$ [rad] | $M_i$ |
|---|---|---|---|
| 0 | 0 | 0 | 0 |
| 1 | 158.5 | 0.932 | 6 |
| 2 | 306.7 | 0.484 | 12 |
| 3 | 456.0 | 0.329 | 18 |
| 4 | 606.8 | 0.25 | 24 |
| 5 | 759.9 | 0.203 | 30 |
| 6 | 915.7 | 0.172 | 35 |
| 7 | 1075 | 0.15 | 41 |
| 8 | 1239 | 0.134 | 46 |
| 9 | 1408 | 0.122 | 50 |
| 10 | 1583 | 0.113 | 54 |
| 11 | 1767 | 0.107 | 58 |
| 12 | 1961 | 0.102 | 60 |
| 13 | 2169 | 0.1 | 62 |
| 14 | 2399 | 0.101 | 61 |
| 15 | 2671 | 0.114 | 54 |
| 16 | 2898 | 0.051 | 121 |
| 17 | 3051 | 0.052 | 120 |
| 18 | 3216 | 0.053 | 117 |
| 19 | 3393 | 0.055 | 114 |
| 20 | 3589 | 0.057 | 109 |
| 21 | 3811 | 0.063 | 99 |
| 22 | 4107 | 0.086 | 72 |
| 23 | 4366 | 0.037 | 166 |
| 24 | 4538 | 0.04 | 157 |
| 25 | 4730 | 0.043 | 145 |
| 26 | 4955 | 0.05 | 125 |

## S5. Python and MATLAB scripts for computing the spatial coordinates of nanorods

Following the theoretical framework detailed in Section S3, this section provides both Python and MATLAB scripts to compute the precise spatial coordinates of all nanorods within an aperiodic metalens. The algorithms output the coordinates in both polar ($r$, $\theta$) and Cartesian ($x$, $y$) systems based on user-defined optical parameters, including the design wavelength $\lambda$, focal length $f$, and the total number of concentric rings $N$. Furthermore, the coordinate generation depends on the specific period profile function, $p(r)$, which is tailored to the selected meta-atom material (e.g., silicon or $TiO_2$) and is derived from the exponential fitting at the design wavelength. In addition to computing and plotting the complete 2D layout to visualize the metalens pattern, the scripts calculate key geometrical metrics, such as the total aperture diameter. Finally, to prevent numerical convergence issues in the root-finding algorithm—particularly when solutions fall exactly on the boundaries between Fresnel zones—the iterative method robustly identifies roots by searching for a sign change instead of a strict zero crossing of the objective function.

## Python code used to compute the spatial coordinates of nanorods

```
# ============================================================
# Aperiodic Metalens Coordinate Generator
#
# Based on Eqs. (S18)-(S21) or (5)-(7)
# "Aperiodic metalenses with identical nanocylinders: positional phase control and passive chromatic aberration suppression "
# ============================================================
#                Python
# ------------------------------------------------------------
# Notes:
# 1) Python indexing starts at 0
# 2) Numerical root solving uses scipy root_scalar()
#    using Brent's method for robust scalar root solving.
# 3) NumPy arrays are used for efficient vectorized operations.
# ============================================================

import numpy as np
import matplotlib.pyplot as plt
from scipy.optimize import root_scalar

# ============================================================
# PARAMETERS
# ============================================================

N = 17                  # Number of rings
wavelength = 660        # Design wavelength [nm]
f = 9900                # Focal length [nm]

# First ring radius
p0 = 150                # Minimum period [nm]
rho0 = p0 / np.sqrt(np.pi)

# ============================================================
# PERIOD vs PHASE FIT
# Eq. (S11)
# Coefficients for Silicon pillars
# (n = 3.835, k = 0.016 at λ = 660 nm)
# ============================================================

a = 0.0983
b = -1.1576
c = 147.4215
d = -0.0961

# ============================================================
# HYPERBOLIC PHASE PROFILE
# Eq. (S13) or (4)
# ============================================================

def phi(r: np.ndarray | float) -> np.ndarray | float:
    """
    Hyperbolic phase profile.
    """
    return -(2 * np.pi / wavelength) * (np.sqrt(r**2 + f**2) - f)

# ============================================================
# 2π WRAPPED PHASE
# ============================================================

def phi_m(r: np.ndarray | float) -> np.ndarray | float:
    """
    Wrapped phase profile.
    """

    wrapped = np.mod(phi(r), -2 * np.pi)
```

```python
    if np.isscalar(wrapped):

        if wrapped > 0:
            wrapped -= 2 * np.pi

    else:

        wrapped = np.where(wrapped > 0, wrapped - 2 * np.pi, wrapped)

    return wrapped

# ============================================================
# PERIOD PROFILE p(r)
# Eq. (S12)
# ============================================================

def p(r: np.ndarray | float) -> np.ndarray | float:
    """
    Period profile obtained from exponential fitting.
    """

    return ( a * np.exp(b * phi_m(r)) + c * np.exp(d * phi_m(r)) )

# ============================================================
# PREALLOCATE ARRAYS
# ============================================================

rho = np.zeros(N + 1)
r = np.zeros(N)
delta_theta = np.zeros(N)
M = np.zeros(N, dtype=int)

# Initial condition
rho[0] = rho0

# ============================================================
# SOLVING Eq. (S18)
#
# rho_i - rho_(i-1) - p(r_i) = 0
#
# where:
#
# r_i = (rho_i + rho_(i-1))/2
# ============================================================

for i in range(N):

    rho_prev = rho[i]

    # ------------------------------------------------------
    # Nonlinear equation
    # ------------------------------------------------------

    def equation(x):

        return (  x - rho_prev - p(0.5 * (x + rho_prev))   )

    # ------------------------------------------------------
    # Root search interval
    # ------------------------------------------------------

    lower = rho_prev + 1.0
    upper = rho_prev + 500.0

    sol = root_scalar(equation, bracket=[lower, upper],method='brentq')
```

```
    rho[i + 1] = sol.root

    # --------------------------------------------------------
    # Eq. (S19)
    # --------------------------------------------------------

    r[i] = 0.5 * (rho[i + 1] + rho[i])

    # --------------------------------------------------------
    # Eq. (S20)
    # --------------------------------------------------------

    delta_theta[i] = p(r[i]) / r[i]

    # --------------------------------------------------------
    # Number of angular segments
    # --------------------------------------------------------

    m_i = (2 * np.pi) / delta_theta[i]

    M[i] = int(np.floor(m_i - 0.65))

# ============================================================
# ANGULAR COORDINATES
# Eq. (S21)
# ============================================================

theta = []

for i in range(N):

    j = np.arange(M[i] + 1)

    theta_i = j * delta_theta[i]

    theta.append(theta_i)

# ============================================================
# CONVERT POLAR COORDINATES TO CARTESIAN
# ============================================================

X = []
Y = []

for i in range(N):

    xi = r[i] * np.cos(theta[i])
    yi = r[i] * np.sin(theta[i])

    X.extend(xi)
    Y.extend(yi)

# Convert to NumPy arrays
X = np.array(X)
Y = np.array(Y)

# ============================================================
# PLOT
# ============================================================

plt.figure(figsize=(8, 8))

W = 15  # Visual width of meta-atoms

plt.scatter( X, Y, s=W, c='red')

plt.axis('equal')
```

```
plt.grid(True)

plt.xlabel('x [nm]')
plt.ylabel('y [nm]')

plt.title('Aperiodic Metalens')

plt.show()

# ============================================================
# METALENS DIAMETER
# ============================================================

D = 2 * r[-1] + p(r[-1])

print('\n')
print(f'Metalens diameter = {D / 1000:.3f} um')

# ============================================================
# DISPLAY NUMERICAL DATA
# ============================================================

print('\n')
print('Radial coordinates r_i [nm]')

# MATLAB-style scientific notation display
np.set_printoptions( precision=4, suppress=False)

print(r / 1000)

print('\n')
print('Angular pitches DeltaTheta_i [rad]')
print(delta_theta)

print('\n')
print('Number of segments M_i')
print(M)
```

**Matlab code used to compute the spatial coordinates of nanorods**

```
%% ==============================================================
%  Aperiodic Metalens Coordinate Generator
%
%  Based on Eqs. (S18)-(S21) or (5)-(7)
%  " Aperiodic metalenses with identical nanocylinders: positional phase control and passive
%   chromatic aberration suppression "
%
%  Some characteristics:
%  1) MATLAB indices start at 1
%  2) Numerical root solving uses fzero()
%  3) Arrays are preallocated for speed
%  4) Cell arrays are used for variable-length theta vectors
% ==============================================================

clear;
clc;
close all;

% =========================
% PARAMETERS
% ==========================

N = 17;                 % Number of rings
lambda = 660;           % Design wavelength [nm]
f = 9900;               % Focal length [nm]

% First ring radius
p0 = 150;               % Minimum period [nm]
rho0 = p0/sqrt(pi);

% =========================
% PERIOD vs PHASE FIT
% Eq. (S11) with coefficients for pillars made of Silicon (n=3.835, k=0.016, for λ=660nm)
% ==========================

a = 0.0983;
b = -1.1576;
c = 147.4215;
d = -0.0961;

% =========================
% Hyperbolic phase profile
% Eq. (S13) or (4)
% ==========================

phi = @(r) -(2*pi/lambda) .* (sqrt(r.^2 + f^2) - f);

% 2pi wrapped phase
phi_m = @(r) mod(phi(r), -2*pi);

% =========================
% Period profile p(r)
% Eq. (S12)
% ==========================

p = @(r) a*exp(b*phi_m(r)) + c*exp(d*phi_m(r));

% ==============================================================
% SOLVING Eq. (S18)
%
% rho_i - rho_(i-1) - p(r_i) = 0
%
% where:
% r_i = (rho_i + rho_(i-1))/2
% ==============================================================
```

```
% Preallocate arrays
rho = zeros(1,N+1);
r   = zeros(1,N);
DeltaTheta = zeros(1,N);
M = zeros(1,N);

% Initial condition
rho(1) = rho0;

% =========================
% Recursive solution
% ==========================

for i = 1:N

    rho_prev = rho(i);

    % Nonlinear equation:
    % rho_i - rho_(i-1) - p[(rho_i+rho_(i-1))/2] = 0

    fun = @(x) x - rho_prev - p(0.5*(x + rho_prev));

    % Initial guess
    x0 = rho_prev + p(rho_prev);

    % Solve nonlinear equation
    rho(i+1) = fzero(fun, x0);

    % Eq. (S19)
    r(i) = 0.5*(rho(i+1) + rho(i));

    % Eq. (S20)
    DeltaTheta(i) = p(r(i))/r(i);

    % Number of angular segments

    m_i = (2*pi)/DeltaTheta(i);

    M(i) = floor(m_i - 0.65);

end

% ==============================================================
% Angular coordinates
% Eq. (S21)
%
% theta_ij = j * DeltaTheta_i
% ==============================================================

theta = cell(1,N);

for i = 1:N

    j = 0:M(i);

    theta{i} = j .* DeltaTheta(i);

end

% ==============================================================
% Convert polar coordinates to Cartesian
% ==============================================================

X = [];
Y = [];
```

```
for i = 1:N

    xi = r(i) .* cos(theta{i});
    yi = r(i) .* sin(theta{i});

    X = [X xi];
    Y = [Y yi];

end

% ============================================================
% PLOT
% ============================================================

figure;
W = 15   % visual width of metal-atoms
scatter(X,Y,W,'r','filled');

axis equal;
grid on;

xlabel('x [nm]');
ylabel('y [nm]');

title('Aperiodic Metalens');

% ============================================================
% METALENS DIAMETER
% ============================================================

D = 2*r(N) + p(r(N));

fprintf('\n');
fprintf('Metalens diameter = %.3f um\n', D/1000);

% ============================================================
% DISPLAY NUMERICAL DATA
% ============================================================

disp(' ');
disp('Radial coordinates r_i [nm]');
disp(r.');

disp(' ');
disp('Angular pitches DeltaTheta_i [rad]');
disp(DeltaTheta.');

disp(' ');
disp('Number of segments M_i');
disp(M.');
```

## S6. Comparison with equal-height meta-atoms for moderate-NA metalenses

This section evaluates the optical performance of the aperiodic and conventional metalenses when both are constrained to an identical nanocylinder height, H. This analysis isolates the effect of spatial distribution by removing volume-dependent absorption discrepancies. The comparison is performed for the metalenses from Section 4.1, but here both architectures utilize an identical height of H=800 nm. All other parameters remain identical to those described in Section 4.1.

Figure S3 shows the phase (Φ) and transmission (T) for the Si conventional meta-atoms at H=800 nm, plotted as a function of the nanocylinder radius. These plots are computed using the same parameters as in Section 4.1, only changing H.

Figure S4 presents the broadband focusing performance over the 120 nm bandwidth (600–720 nm) centered at the design wavelength. The data reveal that, under equal height constraints, both architectures yield similar overall focusing efficiencies. The aperiodic metalens produces a tighter focal spot at the design wavelength. Specifically, it achieves a FWHM of 0.815 µm (only 1.4% larger than its theoretical diffraction limit of 0.804 µm). In contrast, the conventional metalens produces a wider focal spot with a FWHM of 0.870 µm, approximately 12% above the diffraction limit.

The chromatic performance, shown in Figs. S4(a) and S4(b), visually confirms a superior chromatic focal-shift behavior for the aperiodic architecture. Quantitatively, the longitudinal chromatic shift exhibits an inverse trend with respect to wavelength: the focal length increases with wavelength in the conventional metalens, but decreases in the aperiodic design (Fig. S4(c)). The RMSE of the focal length variation is significantly higher for the conventional metalens (1.6 µm) compared to the aperiodic metalens (0.8 µm), a reduction of 48%.

Regarding efficiency, the aperiodic design exhibits higher average focusing efficiency (35%) than the conventional design (33.5%); however, the conventional design exhibits a sharp peak at the design wavelength that rapidly decays, whereas the aperiodic metalens delivers a highly uniform focusing efficiency that even slightly increases at longer wavelengths.

This inverse chromatic behavior is also evident in the focusing resolution (Fig. S4(d)). The FWHM broadens with wavelength for the conventional metalens, whereas it narrows for the aperiodic metalens. Consequently, the focal spot of the aperiodic metalens consistently sharpens across the spectrum as wavelength increases. Although the average spot width is similar for both architectures (aperiodic 850 nm, and conventional 865 nm), the aperiodic design exhibits a smaller FWHM variation, remaining nearly constant between 650–720 nm. Additionally, as indicated by the white dotted lines in Figs. S4(a) and (b), a common focal plane containing all focal points within the depth of focus can be easily identified for the aperiodic metalens; such a broadband focal plane does not exist within the depth of field of the conventional metalens.

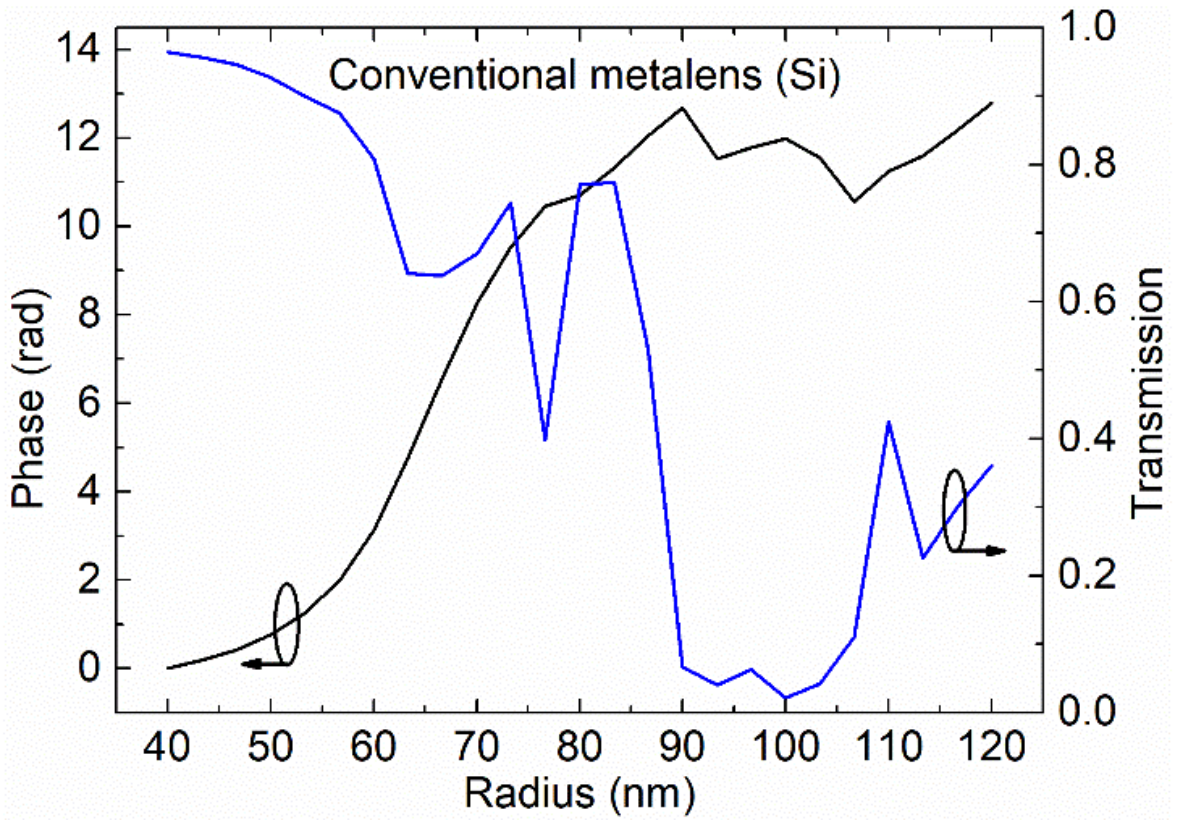

Fig. S3. Phase (left axis) and transmission (right axis) as a function of the radius of the Si meta-atoms with a height of (H=800) nm. These plots are computed at the design wavelength (660 nm), using as refractive index and extinction parameters n=3.835 and k=0.016, for a fixed period p=300 nm. The meta-atom is a square unit cell containing a single nanocylinder (composed of Si) on a fused silica ($SiO_2$) substrate.

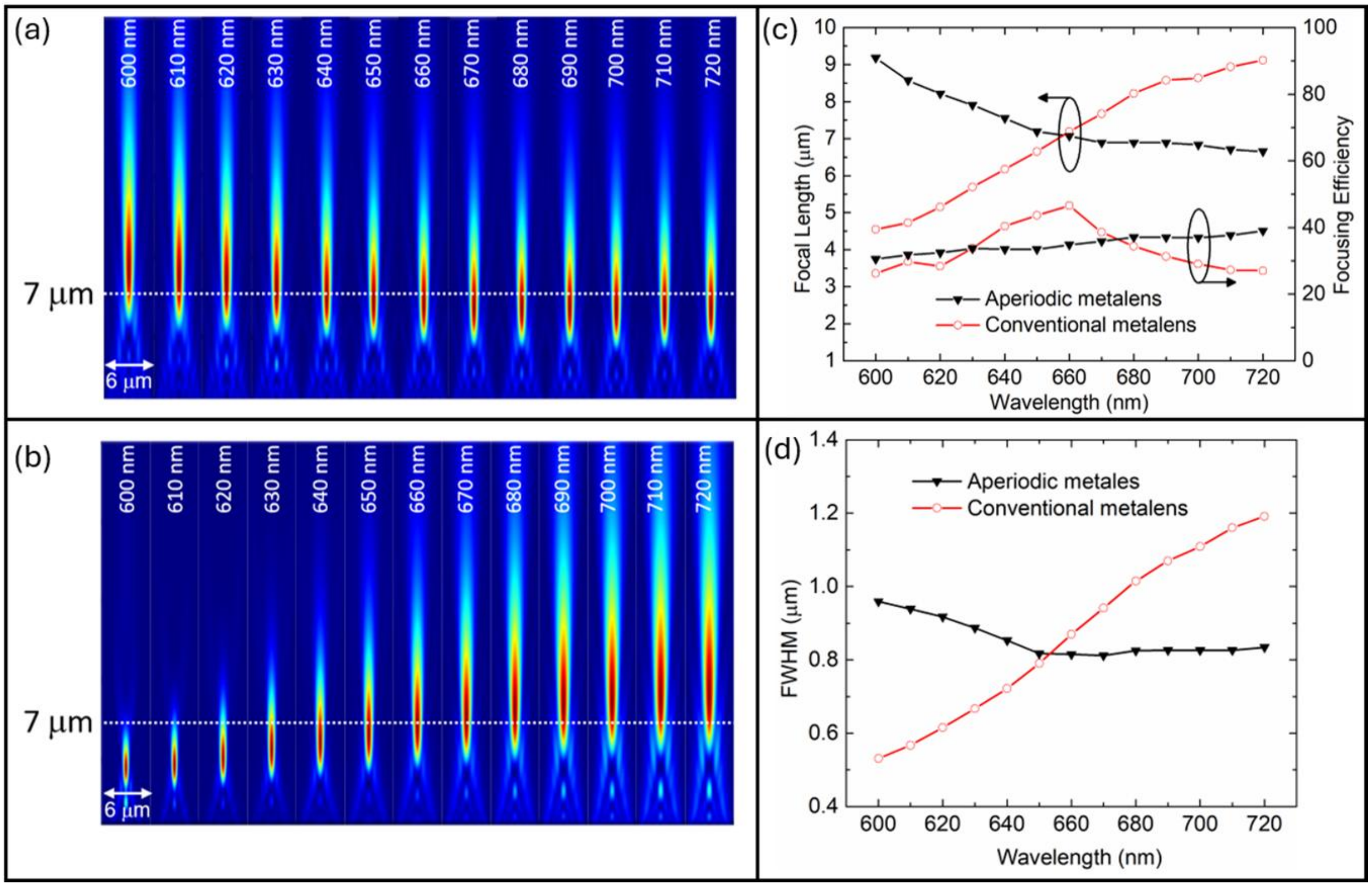

Fig. S4. Chromatic focusing performance of metalenses composed of Si meta-atoms with equal heights and matched unit-cell fill factors. Longitudinal intensity distributions (*xz*-plane) simulated across the 600–720 nm spectral range for (a) the aperiodic metalens, and (b) the conventional size-variant design, illustrating the chromatic focal shift. (c) Quantitative comparison of focal length and focusing efficiency as a function of wavelength. (d) Full Width at Half Maximum (FWHM) of the focal spot versus wavelength. The metalenses analyzed here correspond to the monochromatic designs presented in Fig. 4, which were designed for $\lambda_d$=660 nm, and are evaluated here under broadband illumination. In this case, the meta-atoms have the same height H=800 nm, and identical unit-cell fill factors, enabling a direct comparison of their chromatic focusing performance.

## S7. Depth of focus (DoF) spectral variation for both conventional and aperiodic designs

A large DoF is generally the preferred option if the goal is to maximize achromatic performance over a broad spectral range or if the system must tolerate optical misalignments. However, this comes at the cost of lower spatial resolution; a large DoF intrinsically limits the metalens's ability to resolve small details. Conversely, a small DoF is ideal when the absolute priority is high spatial resolution or precise discrimination between different depth planes. Therefore, the choice of DoF is highly application-dependent. For broadband photography or computer vision, where the goal is to keep the entire visible spectrum in focus simultaneously, a large DoF is preferable. Conversely, for biological microscopy or industrial inspection, where the goal is to resolve nanoscale structures using specific wavelengths, a small DoF is required.

Figure S5 illustrates the DoF as a function of wavelength for both aperiodic and conventional metalenses across the four cases analyzed in Section 4. These include $TiO_2$ meta-atoms (moderate NA) and Si meta-atoms for moderate-NA (17 rings), high-NA (26 rings), and large-diameter (90 rings) metalenses. The DoF is defined as the axial range over which the on-axis intensity remains above 80% of its wavelength-dependent maximum.

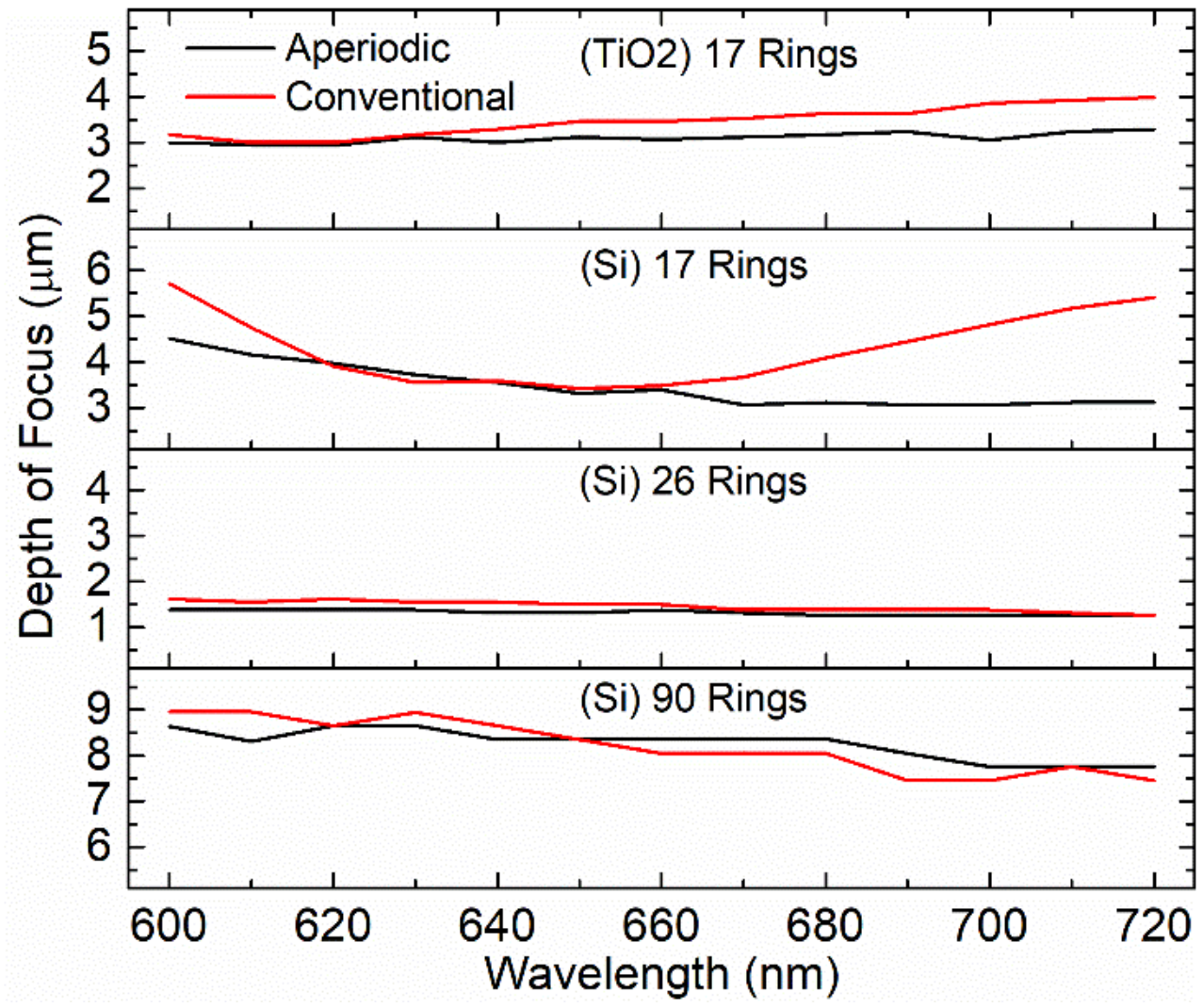

Fig. S5. Depth of focus (DoF) as a function of wavelength for aperiodic (black) and conventional (red) metalenses. Results are shown for $TiO_2$ meta-atoms (17 rings) and Si meta-atoms (17, 26, and 90 rings). The depth of focus is defined as the axial range over which the on-axis intensity remains above 80% of its wavelength-dependent maximum.

## S8. Key performance metrics for both conventional and aperiodic designs

The following Tables S3-S6 summarize the key performance metrics extracted from the FDTD simulations for both the aperiodic and conventional metalenses across the configurations analyzed in Section 4. These data are provided to support the quantitative validation of the proposed physical concept.

Table S3. Comparative Performance Metric for the moderate NA designs. Variation (RMSE) average values are calculated across the 600–720 nm spectral range.

| Parameter | Conventional Metalens | Aperiodic Metalens |
|---|---|---|
| Focal Length Variation (RMSE) | 1.4 μm | **0.8 μm** |
| FWHM @ 660 nm | 0.871 μm | **0.815 μm** |
| Average FWHM of Focal Spot | 1.000 μm | **0.850 μm** |
| Peak Focusing Efficiency | **50%** | 40% |
| Efficiency Variation (RMSE) | 12.3% | **2.5%** |
| Average Efficiency | **39.35%** | 35% |
| Fabrication Requirement | Multiple diameters | **Single diameter** |

Table S4. Comparative Performance Metrics for the high-NA designs. Variation (RMSE) average values are calculated across the 600–720 nm spectral range.

| Parameter | Conventional Metalens | Aperiodic Metalens |
|---|---|---|
| Focal Length Variation (RMSE) | **0.5 μm** | 0.56 μm |
| FWHM @ 660 nm | 0.593 μm | **0.549 μm** |
| Average FWHM of Focal Spot | 597 nm | **557 nm** |
| Peak Focusing Efficiency | **39.6 %** | 37.3 % |
| Efficiency Variation (RMSE) | 12.54 % | **8.82 %** |
| Average Efficiency | 29.6 % | 29.6 % |
| Fabrication Requirement | Multiple diameters | **Single diameter** |

Table S5. Comparative Performance Metrics for the large diameter designs. Variation (RMSE) average values are calculated across the 600–720 nm spectral range.

| Parameter | Conventional Metalens | Aperiodic Metalens |
|---|---|---|
| Focal Length Variation (RMSE) | 4.076 μm | **3.805 μm** |
| FWHM @ 660 nm | 1.230 μm | **1.209 μm** |
| Average FWHM of Focal Spot | 1.234 μm | **1.209 μm** |
| Peak Focusing Efficiency | **36.55 %** | 32.63 % |
| Efficiency Variation (RMSE) | 12.47 % | **11.05 %** |
| Average Efficiency | **26.35 %** | 22.22 % |
| Fabrication Requirement | Multiple diameters | **Single diameter** |

Table S6. Comparative Performance Metric for the $TiO_2$ designs. Variation (RMSE) average values are calculated across the 600–720 nm spectral range.

| Parameter | Conventional Metalens | Aperiodic Metalens |
|---|---|---|
| Focal Length Variation (RMSE) | 0.253 μm | **0.087 μm** |
| FWHM @ 660 nm | 0.877 μm | **0.821 μm** |
| Average FWHM of Focal Spot | 0.871 μm | **0.818 μm** |
| Peak Focusing Efficiency | 53% | **58%** |
| Efficiency Variation (RMSE) | **4.32%** | 4.78% |
| Average Efficiency | 46.18% | **53.38%** |
| Fabrication Requirement | Multiple diameters | **Single diameter** |

## S9. Layouts of large aperiodic metalenses

To further illustrate the scalability of our approach, we present two additional layouts for aperiodic metalenses composed of identical silicon pillars. The coordinate-generation scripts provided in Section S5 are highly robust, enabling the efficient solution of the layout equations (Eqs. 5–7) even for large, millimeter-scale apertures. Figure S6 displays the MATLAB-generated spatial distributions for aperiodic metalenses consisting of 200 and 2000 concentric rings.

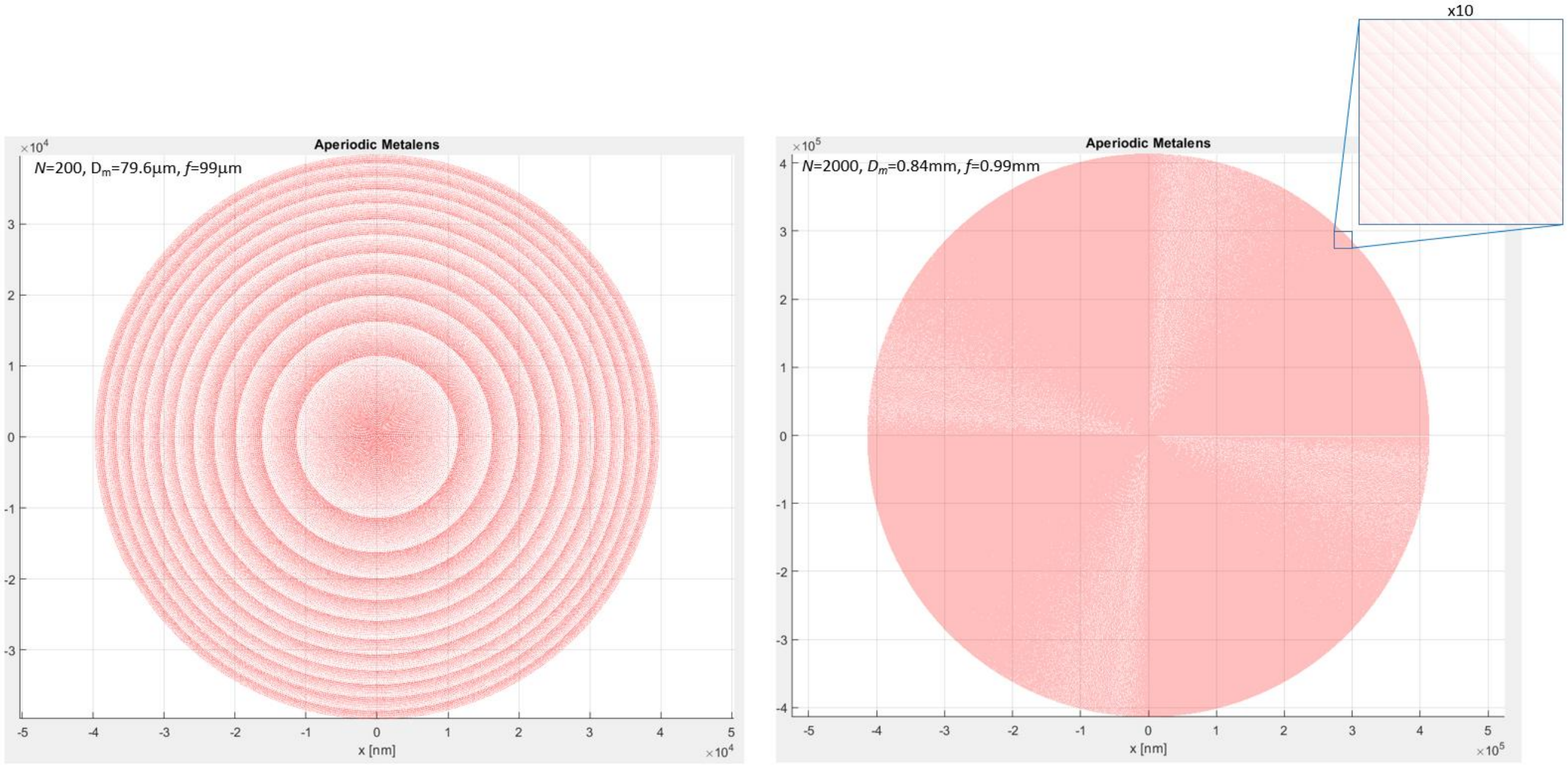

Fig. S6. Layouts of large-scale aperiodic metalenses composed of identical Si nanocylinders. (Left) A metalens design (focal length *f*=99 µm, diameter $D_m$=79.6 µm) consisting of *N*=200 rings. (Right) A millimeter-scale metalens (*f*=0.99 mm, $D_m$=0.84 mm) consisting of *N*=2000 rings.

## S10. Verification of aperiodic unit-cell optical response

This section assesses the optical response of the aperiodic meta-atoms at the densest regions. In particular, we verify the accuracy of the aperiodic meta-atom library by computing not only the phase and transmission responses of an isolated unit cell (as Fig. 2(b)), but also the responses of an array of unit cells via FDTD simulations. The results are displayed in Fig. S7, showing the optical response from 1 unit cell up to a 3x3 array of unit cells, showing that both the phase and transmission response are not significantly affected by the adjacent nanorods.

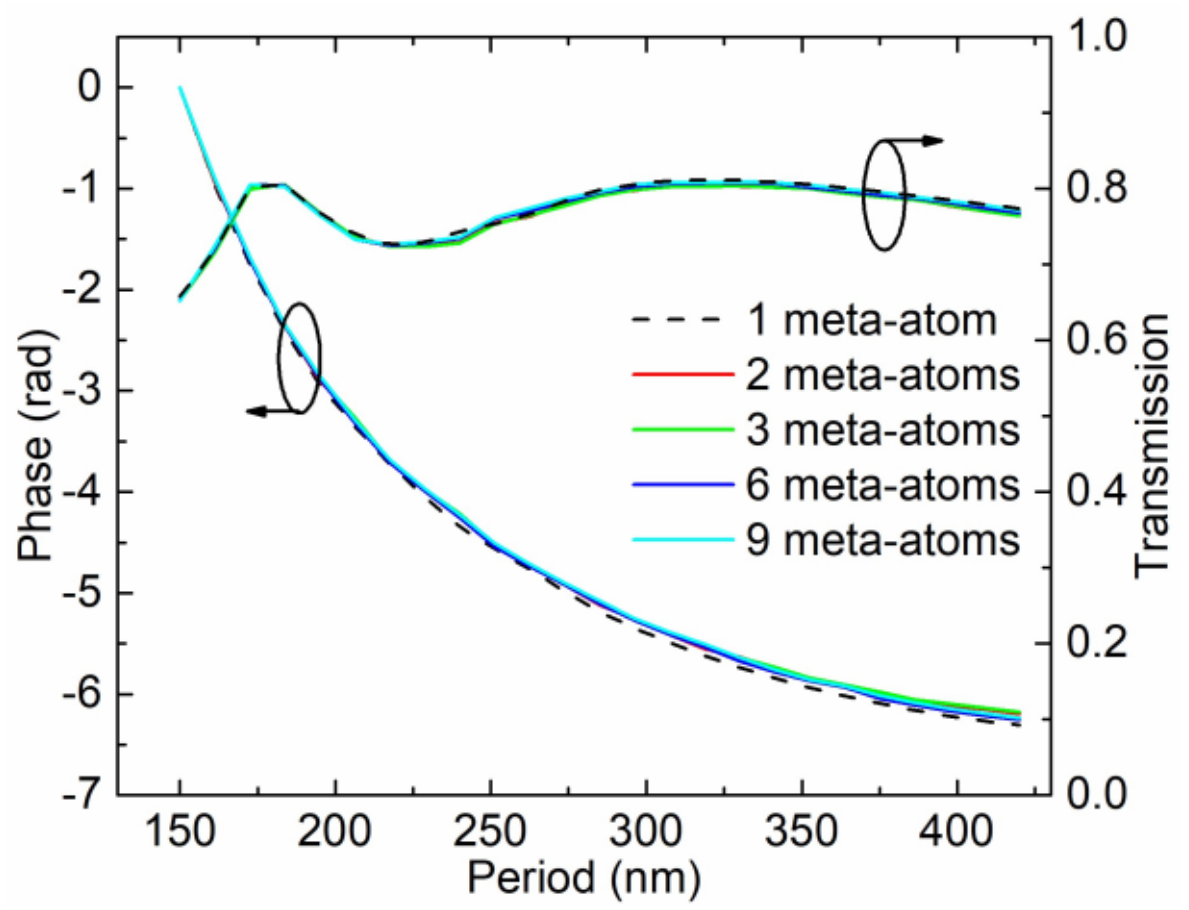

Fig. S7. Phase (left axes) and transmission (right axes) responses as a function of the unit-cell period for an isolated unit-cell (black dashed line) compared with arrays containing 2, 3, 6, and 9 unit-cells, illustrating the effect of neighboring meta-atoms on the optical response.